\documentclass[12pt]{article}%
\usepackage[nosort]{cite}
\usepackage[usenames, dvipsnames]{xcolor}
\usepackage{graphicx}
\usepackage{multicol}
\usepackage{amsfonts}
\usepackage{amssymb}
\usepackage{amsmath}
\usepackage{heck}
\usepackage{afterpage}
\usepackage{setspace}
\usepackage{verbatim}
\usepackage{longtable}
\usepackage{float}
\usepackage{subcaption}
\usepackage{epsfig}
\usepackage{enumerate}
\usepackage{epstopdf}
\usepackage[enableskew, vcentermath]{youngtab}
\usepackage{adjustbox}
\usepackage{multirow}
\usepackage{tikz}
\usepackage[margin=1in]{geometry}
\usepackage{titletoc}
\usepackage{hyperref}
\usepackage{gensymb}
\usepackage{mathtools}%
\usepackage{tikz-cd}
\providecommand{\U}[1]{\protect\rule{.1in}{.1in}}
\pdfoutput=1
\newsavebox{\mysavebox}

\hypersetup{colorlinks,citecolor=black,filecolor=black,linkcolor=black,urlcolor=black}
\usetikzlibrary{decorations.markings}

\numberwithin{equation}{section}

\usetikzlibrary{chains}
\allowdisplaybreaks
\tikzset{node distance=2em, ch/.style={circle,draw,on chain,inner sep=2pt},chj/.style={ch,join},every path/.style={shorten >=4pt,shorten <=4pt},line width=1pt,baseline=-1ex}

\newcommand{\ba}{\begin{eqnarray}}
\newcommand{\ea}{\end{eqnarray}}

\newcommand{\be}{\begin{equation}}
\newcommand{\ee}{\end{equation}}

\tikzstyle{startstop} = [rectangle, rounded corners, minimum width=3cm, minimum height=1cm,text centered, draw=black, fill=blue!10]
\tikzstyle{startstop} = [rectangle, rounded corners, minimum width=3cm, minimum height=1cm,text centered, draw=black, fill=blue!10]
\tikzstyle{io} = [trapezium, trapezium left angle=70, trapezium right angle=110, minimum width=3cm, minimum height=1cm, text centered, draw=black, fill=blue!30]
\tikzstyle{process} = [rectangle, minimum width=3cm, minimum height=1cm, text centered, draw=black, fill=orange!30]
\tikzstyle{decision} = [diamond, minimum width=3cm, minimum height=1cm, text centered, draw=black, fill=green!30]
\tikzstyle{arrow} = [thick,->,>=stealth]
\tikzset{->-/.style={decoration={
  markings,
  mark=at position #1 with {\arrow[scale=2.4]{>}}},postaction={decorate}}}
\makeatletter \@addtoreset{equation}{section} \makeatother

\begin{document}

\date{June 2019}

\title{T-Branes and $G_2$ Backgrounds}

\institution{PENN}{\centerline{${}^{1}$Department of Physics and Astronomy, University of Pennsylvania, Philadelphia, PA 19104, USA}}

\institution{PENNMATH}{\centerline{${}^{2}$Department of Mathematics, University of Pennsylvania, Philadelphia, PA 19104, USA}}

\institution{MARIBOR}{\centerline{${}^{3}$Center for Applied Mathematics and Theoretical Physics, University of Maribor, Maribor, Slovenia}}

\authors{Rodrigo Barbosa\worksat{\PENNMATH}\footnote{e-mail: {\tt barbosa@sas.upenn.edu}},
Mirjam Cveti\v{c}\worksat{\PENN, \PENNMATH, \MARIBOR}\footnote{e-mail: {\tt cvetic@physics.upenn.edu}},
Jonathan J. Heckman\worksat{\PENN}\footnote{e-mail: {\tt jheckman@sas.upenn.edu}},\\[4mm]
Craig Lawrie\worksat{\PENN}\footnote{e-mail: {\tt craig.lawrie1729@gmail.com}},
Ethan Torres\worksat{\PENN}\footnote{e-mail: {\tt emtorres@sas.upenn.edu}},
and Gianluca Zoccarato\worksat{\PENN}\footnote{e-mail: {\tt gzoc@sas.upenn.edu}}}

\abstract{Compactification of M- / string theory on manifolds with $G_2$ structure yields a wide variety of
4D and 3D physical theories. We analyze the local geometry of such compactifications as captured by a
gauge theory obtained from a three-manifold of ADE singularities.
Generic gauge theory solutions include a non-trivial gauge field flux as well as normal deformations to the
three-manifold captured by non-commuting matrix coordinates, a signal of T-brane phenomena.
Solutions of the 3D gauge theory on a three-manifold are
given by a deformation of the Hitchin system on a marked Riemann surface
which is fibered over an interval. We present explicit examples of
such backgrounds as well as the profile of the
corresponding zero modes for localized chiral matter. We also
provide a purely algebraic prescription for
characterizing localized matter for such T-brane configurations.
The geometric interpretation of this gauge theory description
provides a generalization of twisted connected sums
with codimension seven singularities at localized regions
of the geometry. It also indicates that geometric codimension six
singularities can sometimes support 4D chiral matter due to physical
structure ``hidden'' in T-branes.}

{\small \texttt{\hfill UPR-1298-T}}

\maketitle

\setcounter{tocdepth}{2}

\tableofcontents


\newpage

\section{Introduction \label{sec:INTRO}}

Manifolds of special holonomy are of great importance in connecting the
higher-dimensional spacetime predicted by string theory to lower-dimensional
physical phenomena. This is because such manifolds admit covariantly constant
spinors, thus allowing the macroscopic dimensions to preserve some amount of
supersymmetry.

Historically, the most widely studied class of examples has centered on type
II\ and heterotic strings compactified on Calabi--Yau threefolds \cite{Candelas:1985en}. This leads to 4D\
vacua with eight and four real supercharges, respectively. Such threefolds also
play a prominent role in the study of F-theory and M-theory backgrounds,
leading respectively to 6D and 5D\ vacua with eight real supercharges.
Compactifications on Calabi--Yau spaces of other dimensions lead to a rich class
of geometries, and correspondingly many novel physical systems in the
macroscopic dimensions. In all these cases, the holomorphic
geometry of the Calabi--Yau allows techniques from
algebraic geometry to be used.

There are, however, other manifolds of special holonomy, most notably those
with $G_{2}$ and $Spin(7)$ structure. For example, compactification of M-theory
on $G_{2}$ and $Spin(7)$ spaces provides a method for generating a broad class
of 4D and 3D $\mathcal{N}=1$ vacua, respectively.\footnote{In F-theory, there has recently been renewed interest in the
use of $Spin(7)$ backgrounds as a way to generate novel models of dark
energy \cite{Heckman:2018mxl,Heckman:2019dsj} (see also \cite{Bonetti:2013fma,Bonetti:2013nka, Berglund:2019pxr, Berglund:2019ctg}).
Though less studied, F-theory on $G_{2}$ backgrounds should also lead to novel 5D vacua \cite{Vafa:1996xn}.}

Despite these attractive features, it has also proven notoriously difficult to generate singular \emph{compact} geometries of direct relevance for physics. In the case of M-theory on a $G_2$ background, realizing a non-abelian ADE gauge group requires a three-manifold of ADE singularities (i.e., codimension four), and realizing 4D chiral matter requires codimension seven singularities. While there are now some techniques available to realize $G_2$ backgrounds with codimension four singularities, it is not entirely clear whether a smooth compact $G_2$ can be continuously deformed to such singular geometries, as necessary for physics.\footnote{Recently, the local model version of this problem has been
solved \cite{barbosa, barbosa2019}. The main result is the construction of a
deformation family of closed $G_2$-structures starting from a given
$G_2$-structure on the total space of a fibration of ADE singularities. In a
nutshell, the deformations are parametrized by certain spectral covers in a local gauge theory (detailed later in
this paper). This result is a $G_2$ analogue of the
well-known correspondence between Calabi--Yau threefolds ALE-fibered over a
Riemann surface $\Sigma$ and Hitchin systems over $\Sigma$
\cite{Diaconescu:2005jw, Diaconescu:2006ry}.}

Once one is given a codimension four ADE singularity, further degenerations at points of the three-manifold should produce the codimension seven singularities\footnote{There are quite a few known examples of local models of $G_2$ spaces with \emph{conical} singularities. A $G_2$-metric on a seven-dimensional cone is equivalent to a nearly-K\"ahler metric on the base of the cone, and these are known to exist on $S^6$, $\mathbb{CP}^3$, $SU(3)/U(1)^2$ and $S^3\times S^3$. It is known \cite{foscolo1, foscolo2, gukovetal, cvetic} that one-parameter families of $G_2$-metrics exist deforming the $G_2$-structure on the cone over $S^3\times S^3$. Analogous $Spin(7)$-metrics were constructed in \cite{Cvetic:2001pga} (see also \cite{Cvetic:2001ye} and the review \cite{Cvetic:2002kn}).} required for 4D chiral matter.\footnote{4D $\mathcal{N} = 1$ globally consistent type IIA compactifications with chiral matter \cite{Cvetic:2001tj,Cvetic:2001nr} and their relation
to M-theory on compact $G_2$ holonomy spaces were studied
in reference \cite{Cvetic:2001kk}.} It is expected that some of the fibers in the deformation family of \cite{barbosa} could acquire
such point-like singularities, and we refer to that paper for a discussion on how that could be achieved from a geometric perspective.

In this approach to singular $G_2$ compactifications, rather than building the global geometry directly, the crucial idea is to use a dual gauge-theoretic description to characterize the appearance of such
codimension seven singularities. In reference \cite{Pantev:2009de}, the partial topological
twist of a six-brane wrapped on a three-manifold $M$ embedded in a $G_{2}$
manifold was studied in some detail, and we shall refer to it as the
\textquotedblleft Pantev--Wijnholt\textquotedblright\ (PW) system. The choice
of 4D vacuum is dictated in the six-brane gauge theory by an adjoint-valued one-form and a
vector bundle. The eigenvalues of this one-form parameterize normal
deformations in the local geometry $T^{\ast}M$, and this leads to a
natural spectral cover description. Localized matter in
this setup is obtained by allowing the one-form to vanish at various
locations. In \cite{Pantev:2009de} this was used to analyze codimension seven singularities,
and in reference \cite{Braun:2018vhk} this analysis was greatly developed and also
extended to the case of codimension six singularities, i.e. non-chiral matter.
As argued in \cite{Braun:2018vhk} a potentially appealing feature of these
codimension six singularities is that they provide a way to possibly connect
to one of the (few)\ methods available for building $G_{2}$ manifolds via
twisted connected sums, using Calabi--Yau threefolds as building blocks \cite{MR2024648,MR3109862,MR3369307}. The physics of M-theory compactified on such compact TCS $G_2$ manifolds has been studied in \cite{Halverson:2014tya,Halverson:2015vta,Braun:2016igl,Guio:2017zfn,Braun:2017ryx,Braun:2017uku,
Braun:2017csz,Braun:2018fdp,Fiset:2018huv,Acharya:2018nbo,Braun:2018vhk}.

In the TCS construction of $G_{2}$ backgrounds \cite{MR2024648, MR3369307}, one first
begins with a pair $(M_1,M_2)$ of $G_2$-manifolds of the form $M_i=X_i\times S^1_{i,\text{ext}}$, where $S^1_{i,\text{ext}}$ is a circle we call ``external'' and $X_i$ is an \emph{asymptotically cylindrical} Calabi--Yau threefold. This means that outside of a compact region, the Calabi--Yau metric looks like a $K3$ surface times a cylinder: $X_i \stackrel{\text{asymp.}}{\cong} K3_i\times \mathbb{R} \times S^1_{i,\text{int}}$. One then glues $S^1_{1,\text{ext}}$ and $S^1_{2,\text{int}}$ using a hyperk\"ahler rotation (and similarly for the other two circles) and shows that the resulting compact smooth manifold admits a full $G_{2}$
metric. Although all known examples result in smooth total spaces, it is
reasonable to expect that degenerations in the Calabi--Yau building blocks will
provide a way to generate the long sought for codimension six and seven
singularities in compact models.

\begin{figure}[t!]
\begin{center}
\includegraphics[trim={0cm 2.5cm 0cm 0cm},clip,scale=0.5]{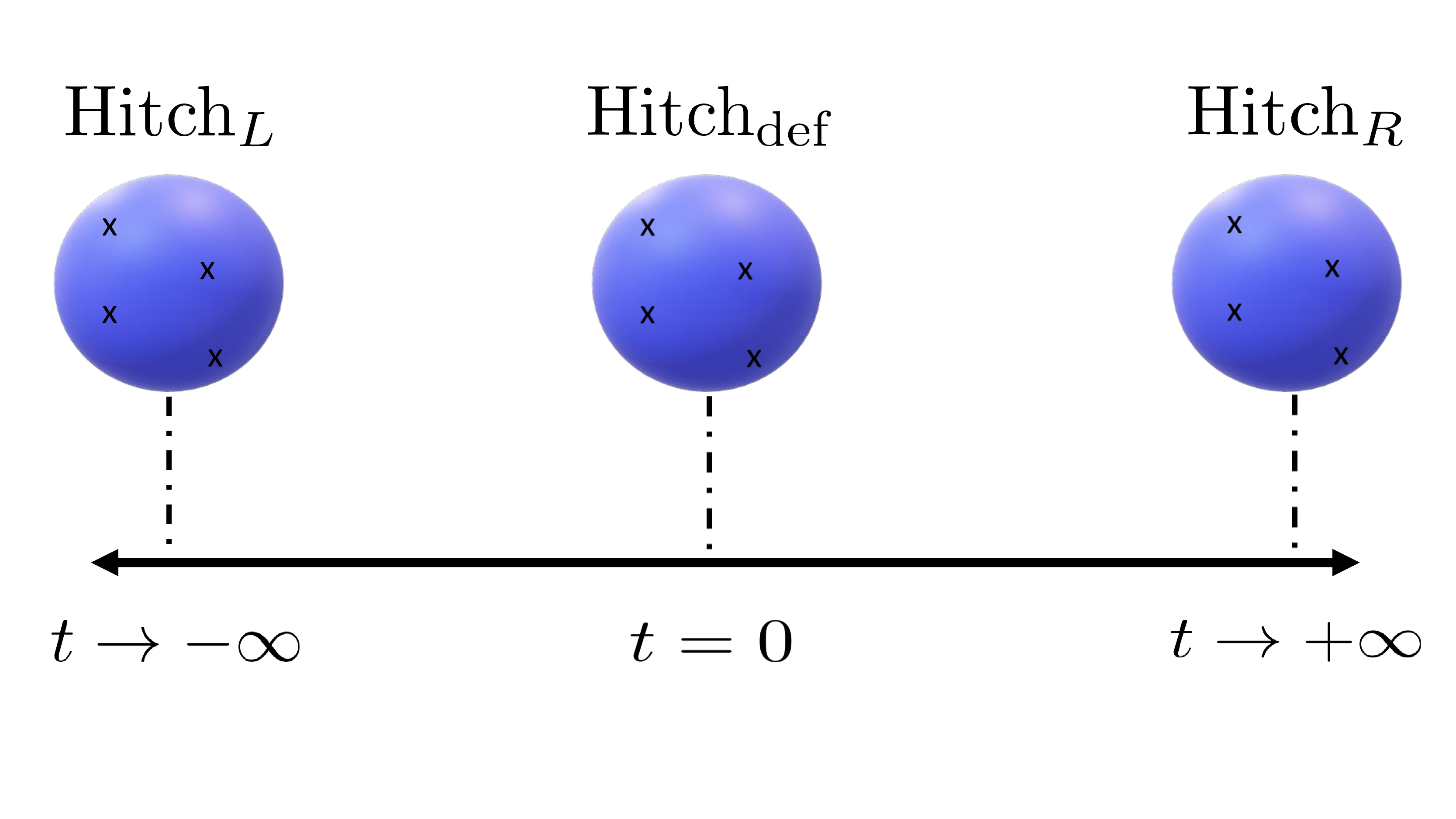}
\end{center}
\caption{The local $G_2$ background of a three-manifold of ADE
singularities is characterized by gauge theory on a three-manifold with corresponding ADE gauge group.
In a local patch, this can be described by a Riemann surface $\Sigma$ with marked points
fibered over an interval. In a suitable scaling limit of the metric, this can be
viewed as a deformation of the Hitchin system over $\Sigma$ which asymptotes to
solutions to the Hitchin system. See figure \ref{fig:G2FIBER} for a depiction of the geometry
associated with this local gauge theory.}
\label{fig:HITCHFIBER}
\end{figure}

\begin{figure}[t!]
\begin{center}
\includegraphics[trim={0cm 2.5cm 0cm 0cm},clip,scale=0.5]{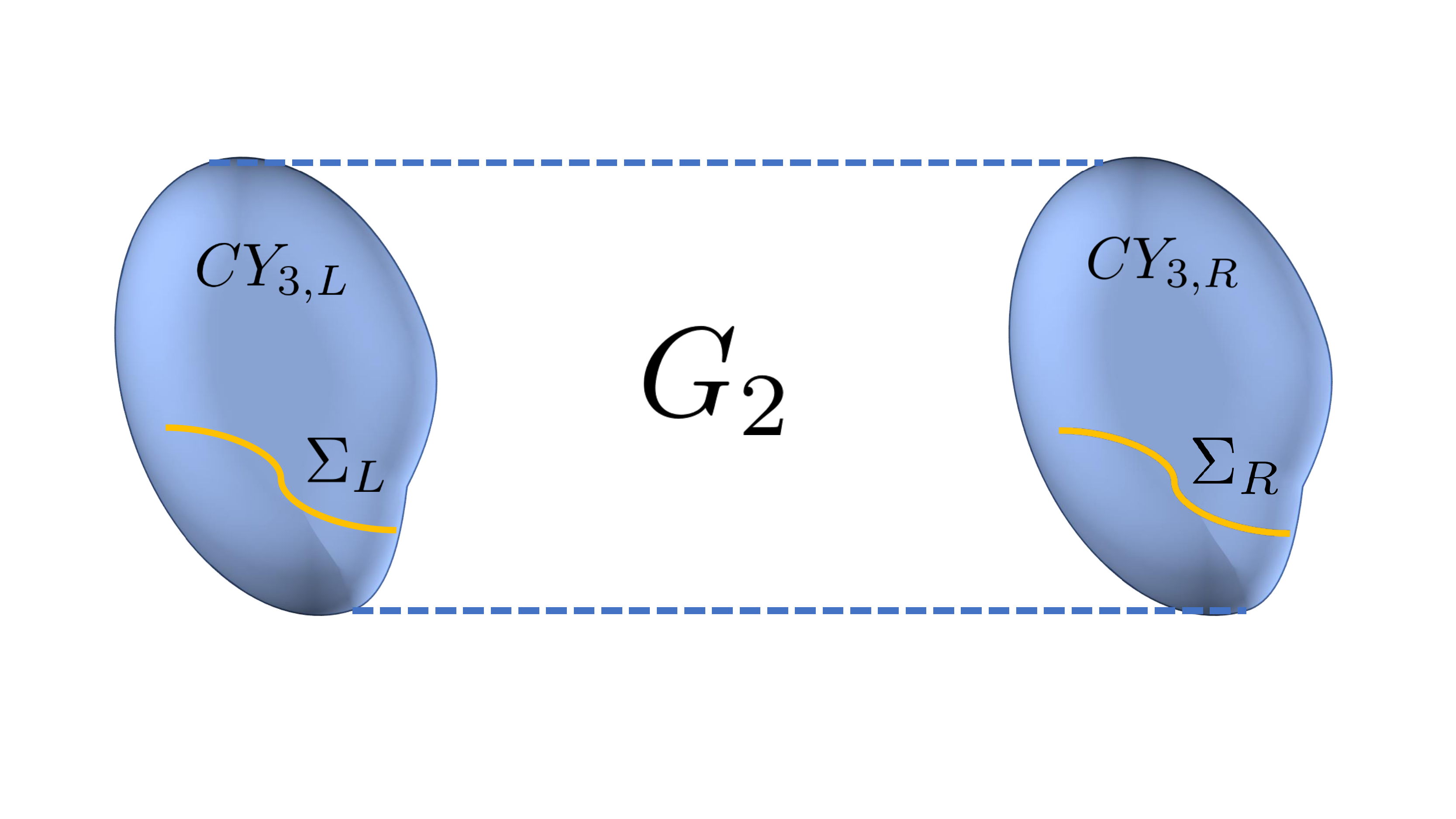}
\end{center}
\caption{
	The local gauge theory analysis allows us to build up local $G_2$ backgrounds which asymptote to Calabi--Yau threefolds at
two boundaries. The local $G_2$ background includes a three-manifold which is itself a Riemann surface $\Sigma$ fibered over
an interval. This Riemann surface embeds in the boundary Calabi--Yau spaces.
See figure \ref{fig:HITCHFIBER} for a depiction of the asymptotic behavior of the gauge theory on the three-manifold as a
deformation of the Hitchin system on $\Sigma$.}
\label{fig:G2FIBER}
\end{figure}

From this perspective, one might ask whether this is the most general
starting point one can entertain for realizing localized matter in $G_{2}$
compactifications. One important clue comes from the structure of local $G_2$ backgrounds
in the presence of non-zero fluxes. This leads to manifolds with structure group $G_2$,
and these solutions can often be interpreted in terms of a system of lower-dimensional
branes localized on subspaces inside the bulk geometry \cite{Cvetic:2001ya, Cvetic:2001zx, Chong:2002yh}.
It is thus natural to ask about fluxed solutions with lower-dimensional defects.

In this paper we consider the case of non-abelian fluxes of a six-brane wrapped on a
three-manifold. To accomplish this, we return to the local
gauge theory on a six-brane (see also \cite{Acharya:1998pm, Acharya:2001gy, Braun:2018vhk}).
To date, most analyses of localized matter have assumed that the adjoint valued one-form is diagonal,
and that there are no fluxes present in the three-manifold. Here, we shall
relax this assumption and attempt to study a far broader class of
situations. This necessarily means that the components of this one-form will
not commute. We shall refer to this as a T-brane configuration (even though
the matrix components are not upper triangular) since it naturally fits in the
broader scheme of T-brane like phenomena. For earlier work on T-branes, see references
\cite{Aspinwall:1998he, Donagi:2003hh,Cecotti:2009zf,Cecotti:2010bp,Donagi:2011jy,Anderson:2013rka,
Collinucci:2014qfa,Cicoli:2015ylx,Heckman:2016ssk,Collinucci:2016hpz,Bena:2016oqr,
Marchesano:2016cqg,Anderson:2017rpr,Collinucci:2017bwv,Cicoli:2017shd,Marchesano:2017kke, Cvetic:2018xaq, 
Heckman:2018pqx, Apruzzi:2018xkw,Carta:2018qke,Marchesano:2019azf}.

Locally modelling the three-manifold as a Riemann surface fibered over an
interval, we show that for each smooth fiber, the gauge theory on the Riemann
surface is described by a mild deformation of Hitchin's system on a complex
curve (see figure \ref{fig:HITCHFIBER}). Since the local Hitchin system directly describes a local Calabi--Yau
geometry (see e.g. \cite{Donagi:1995cf,Diaconescu:2005jw,Diaconescu:2006ry,Anderson:2013rka,Anderson:2017rpr}), we obtain
a local deformation of a TCS-like construction which can
be interpreted as building up a local $G_{2}$ background (see figure \ref{fig:G2FIBER}).

An important feature of these structures is the appearance of holomorphic geometry as a
guide in constructing these local $G_2$ backgrounds. This points the way to a
method for constructing $G_2$ backgrounds using more general holomorphic building
blocks than those appearing in the classical TCS construction.

Another feature we study in great detail is the resulting localized matter
obtained from such T-brane configurations. We provide examples where we explicitly
determine the profile of localized matter fields in a given background.
This involves solving a second order differential equation. We also
develop algebraic methods for reading off the appearance of localized zero modes
by determining the local ring structure of trapped matter. This is similar in spirit to the analysis of
localized zero modes in T-brane configurations carried out in references
\cite{Cecotti:2009zf,Cecotti:2010bp}.

One of the outcomes of this analysis is that it also provides evidence for the
existence of localized matter field configurations which would be
\textquotedblleft invisible\textquotedblright\ to the bulk $G_{2}$ geometry
since they originate from degenerate spectral equations. Instead, they would
be fully characterized by allow limiting behavior in the four-form $G$-flux of
the M-theory background. We present explicit examples exhibiting this behavior.
A canonical example is the ``standard embedding'' namely we embed the spin connection in the
gauge bundle, taking our holomorphic vector bundle to be the tangent bundle on $T^{\ast}M$
dimensionally reduced (after a Fourier-Mukai transform / formally three T-dualities) to the
three-manifold. Related examples show up in a
number of other T-brane constructions (see e.g. \cite{Aspinwall:1998he,Anderson:2013rka}).

The rest of this paper is organized as follows. We begin in section
\ref{sec:GAUGE}\ by discussing the PW system, and its relation to
geometric engineering. Next, in section
\ref{sec:BKGND}\ we show how to build examples of solutions to the PW
system in cases with non-zero gauge field flux. In section \ref{sec:MATTER}
we present some general methods for analyzing zero modes
in such backgrounds, and then proceed to determine the localized matter field
wave functions by explicitly solving the corresponding partial differential equations.
Section \ref{sec:ALGEBRAIC} presents a conjectural proposal for how to algebraically determine
the localized zero mode content in fluxed solutions. We present our
conclusions in section \ref{sec:CONC}.

\section{Six-Brane Gauge Theory on a Three-Manifold \label{sec:GAUGE}}

In this section we discuss the gauge theory of a six-brane wrapped on a three-manifold,
as obtained from M-theory on a $G_{2}$ background. Geometrically, we engineer
this gauge theory from a three-manifold of ADE\ singularities with
corresponding ADE\ gauge group - i.e., the local $G_2$ is a fibration of
ADE singularities over a three-manifold. The terminology follows from the fact that the
\textquotedblleft brane\textquotedblright\ in question is actually a
seven-dimensional supersymmetric gauge theory wrapped on the three-manifold,
namely a six-brane. Supersymmetry is preserved since we assume the three-manifold
is an associative three-cycle of the local $G_2$
manifold.\footnote{Associative three-cycles are three
manifolds on which the associative three-form $\Phi$ of a $G_2$ manifold restricts to the volume form of the three manifold. See for example \cite{mclean} for more details on associative three-cycles and their deformation theory.} It is also in
accord with the type IIA\ string theory description of D6-branes wrapped
on special Lagrangian submanifolds of a Calabi--Yau threefold.
For a recent pedagogical discussion of geometric
engineering, and the relation between localized gauge theories and singular geometry
in the context of string compactification, see the online lectures of reference
\cite{HeckmanLectures}.

In terms of the geometry of the local $G_{2}$ background, we note that we have
an associative three-form $\rho$ which pairs with the three-form
potential $C$ of M-theory to form the complex moduli $C + i\rho$.
Resolving the ADE fibers and performing a corresponding reduction to the
three-manifold, we have a decomposition:%
\begin{equation}
C=A_{\alpha}\wedge\omega^{\alpha} \label{rhoC}\text{, \ \ } \rho=\phi_{\alpha}\wedge\omega^{\alpha}
\end{equation}
where the $\omega^{\alpha}$ are harmonic representatives of $(1,1)$ forms on the resolution of
the local ADE\ singularity, and the $A_{\alpha}$ and $\phi_{\alpha}$ are
one-forms on the three-manifold. Moreover $\alpha=1,\dots ,r $ with $r$ being the rank of the ADE gauge group. One should  think of  this index as labeling the generators of the Cartan subalgebra of the corresponding
ADE\ gauge group. The remaining generators are obtained from M2-branes wrapped
on collapsing two-cycles of the fiber.

As explained in reference \cite{Pantev:2009de} (see also
\cite{Acharya:1998pm,Braun:2018vhk,Blau:1996bx,Bershadsky:1995qy}), the partial twist of a
six-brane with gauge group $G$ on a three-manifold $M$ retains 4D
$\mathcal{N}=1$ supersymmetry in the uncompactified directions. After the twist both the gauge field $A$ and $\phi$ become adjoint-valued one-forms on $M$. They combine into a complexified connection which we write
as:
\begin{equation}
\label{eq:calA}
\mathcal{A}=A+i \phi,
\end{equation}
which should be thought of as the bosonic component of a collection of 4D
$\mathcal{N}=1$ chiral superfields which transform as a one-form on the three
manifold. In our conventions, we take anti-hermitian generators for the Lie algebra
so that $A^{\dag} = - A$ and $\phi^{\dag} = - \phi$. We shall also find it convenient
to absorb the factor of $i$ to define a hermitian Higgs field:
\begin{equation}
\Phi \equiv i \phi.
\end{equation}

We construct various curvatures from the complexified connection $D_{\mathcal{A}}$, its
conjugate $D_{\overline{\mathcal{A}}}$ as obtained from $\overline
{\mathcal{A}}=A-i \phi$ as well the purely real $D_{A}$. In a unitary gauge, we
have $\mathcal{A}^{\dag} = - \overline{\mathcal{A}}$.
Locally these connections may be written as:%
\begin{equation}
D_{\mathcal{A}}=d+\mathcal{A}\text{, \ \ }D_{\overline{\mathcal{A}}%
}=d+\overline{\mathcal{A}}\text{, \ \ }D_{A}=d+A.
\end{equation}
We introduce gauge field strengths:%
\begin{equation}
\mathcal{F}=\left[  D_{\mathcal{A}},D_{\mathcal{A}}\right]  \text{,
\ \ }\mathcal{D}=\left[  D_{\mathcal{A}},D_{\overline{\mathcal{A}}%
}\right]  \text{, \ \ }F=\left[  D_{A},D_{A}\right]  .
\end{equation}
Once written in components the various field strengths are:
\begin{align}
\mathcal{F}_{ij}  &  =\partial_{i}\mathcal{A}_{j}-\partial_{j}\mathcal{A}%
_{i}+\left[  \mathcal{A}_{i},\mathcal{A}_{j}\right] \\
\mathcal{D}_{ij}  &  =\partial_{i}\overline{\mathcal{A}}_{j}-\partial
_{j}\mathcal{A}_{i}+\left[  \mathcal{A}_{i},\overline{\mathcal{A}}_{j}\right]
\\
F_{ij}  &  =\partial_{i}A_{j}-\partial_{j}A_{i}+\left[  A_{i},A_{j}\right]  .
\end{align}
In terms of the original fields $A$ and $\phi$, the complexified field
strengths decompose as:
\begin{align}
\operatorname{Im}\mathcal{F}_{ij}  &  =F_{ij}-[\phi_{i},\phi_{j}]\\
\operatorname{Re}\mathcal{F}_{ij}  &  =(\partial_{i}+A_{i})\cdot\phi
_{j}-(\partial_{j}+A_{j})\cdot\phi_{i}\\
\operatorname{Re}\mathcal{D}_{ij}  &  =-(\partial_{i}+A_{i})\cdot\phi_{j}-(\partial_{j}+A_{j})\cdot\phi_{i}\\
\operatorname{Im}\mathcal{D}_{ij}  &  =F_{ij} + [\phi_i,\phi_j].
\end{align}

We are considering here the equations for the fields supported
on the three-manifold, and thus the the indices $i,j$ run from $1$ to $3$.
Variation of the 7D\ gaugino produces the corresponding conditions to have a 4D
$\mathcal{N}=1$ supersymmetric vacuum. These are conveniently packaged as F-terms and
D-terms, which are respectively metric independent and dependent:%
\begin{equation}
\mathcal{F}_{ij}=0\text{ \ \ and \ \ }g^{ij}\mathcal{D}_{ij}=0,
\label{pweqs}
\end{equation}
in the obvious notation. The moduli space of vacua is then given by two
equivalent presentations:%
\begin{align}
\{\mathcal{F}  &  =0\}/G_{\mathbb{C}}^{\text{gauge}}\\
\{\mathcal{F}  &  =0\text{, \ \ }g^{ij}\mathcal{D}_{ij}=0\}/G_{U}%
^{\text{gauge}}\text{,}%
\end{align}
where here, $G_{\mathbb{C}}^{\text{gauge}}$ refers to complexified gauge
transformations and $G_{U}^{\text{gauge}}$ refers to unitary gauge
transformations. We refer to both presentations of the moduli space as the
defining equations of the Pantev--Wijnholt (PW)
system.\footnote{These equations are in fact part of a
one-parameter family of equations that can obtained by dimensionally reducing
three directions of the six-dimensional Hermitian Yang-Mills equations. This is discussed, for example, in Appendix A of
\cite{Gaiotto:2011nm} to which we refer the interested reader for further details. Denoting the associated parameter by $\zeta\in\mathbb{R}$, then for $\zeta\neq 0$, the set of equations are the PW F-term equations along with a modified D-term which is of PW form for $\zeta=\pm 1$. When $\zeta=0$, these equations are the so-called Extended Bogomolny equations which have an additional adjoint one-form compared to the normal Bogomolny equations. In the presence of a 2d boundary, a Nahm pole-like boundary condition can be imposed for any $\zeta$, which in the PW case would indicate the presence of M5-branes embedded in our gauge theory six-brane. Such configurations would generically induce a coupling to a sector with a non-trivial IR fixed point and for simplicity we leave their study for future work.}

The F-term equations of motion are obtained from the critical points of the Chern-Simons superpotential
for the complexified connection:
\begin{equation}
\mathcal{W}\left[  \mathcal{A}\right]  =\frac{s}{4\pi}\int_{M}\text{Tr}%
_{\mathfrak{g}_{\mathbb{C}}}\left(  \mathcal{A\wedge}d\mathcal{A+}\frac{2}%
{3}\mathcal{A\wedge A\wedge A}\right)  ,
\end{equation}
for $s$ a complex parameter. Four-dimensional $\mathcal{N}=1$ vacua are
labelled by critical points of $\mathcal{W}$
modulo $G_{\mathbb{C}}^{\text{gauge}}$.

More precisely, in the description specified by F-terms modulo $G_{\mathbb{C}}^{\text{gauge}}$ gauge transformations, the appropriate notion of
``stability'' is that we only look at connections with semisimple
monodromy.\footnote{A connection has semisimple monodromies
if the map $\mathcal A : \pi_1(M) \rightarrow G_{\mathbb C}$ gives a semisimple representation of the fundamental group of $M$. This means it is not possible to conjugate the monodromies of $\mathcal A$ to a block triangular form without being able to bring them to a block diagonal form.}
By the Donaldson--Corlette theorem \cite{donaldson, corlette}, these automatically solve the harmonic metric equation, i.e. the D-term. An interesting feature is that for any hermitian
generator of the algebra, the signature of the real matrix $\mathcal{D}%
_{ij}^{B}$ (with $B$ an index in the adjoint) must have at least one $+$ sign
and one $-$ sign each. This is simply to satisfy the D-term constraint. We
note that this is in accord with the \textquotedblleft Hessian
condition\textquotedblright\ of references \cite{Pantev:2009de,Braun:2018vhk} observed in the
special case where the adjoint-valued one-form is diagonal. In this case, the eigenvalues must have, in some basis,
signature $(+,+,-)$ or $(-,-,+)$. The vanishing locus of the Higgs field then specifies a chiral or anti-chiral zero mode.

Another way to study this system is to first consider stable holomorphic vector
bundles on the local Calabi--Yau $T^{\ast}M\equiv X$. These are
described by Hermitian--Yang--Mills (HYM) instantons \cite{duy, duy2}. Taking the
linearization of the HYM equations in a neighborhood of the zero section $M$
then produces the same equations \cite{Pantev:2009de}. Briefly summarizing this approach, we
introduce a $(0,1)$ connection:%
\begin{equation}
\overline{\partial}_{\mathbb{A}}=\overline{\partial}+\mathbb{A}\text{.}%
\end{equation}
The conditions to have a stable holomorphic vector bundle are a F-term and a
D-term:%
\begin{equation}
\mathbb{F}_{(0,2)}=0\text{ \ \ and \ \ }J\wedge J\wedge\mathbb{F}%
_{(1,1)}=0\text{,}%
\end{equation}
with $J$ the K\"{a}hler form on $X$. To make contact with the PW
equations, we consider $\mathbb{A}$ as a complexified gauge field which splits
up as:%
\begin{equation}
\mathbb{A}=A_{i}dx^{i}+\phi_{j}dp^{j}, \label{complexconnection}%
\end{equation}
where $x$ are local coordinates on $M$ and $p$ are local coordinates in
the cotangent direction. The topological twist amounts to making a further
identification $dp^{i}=J_{j}^{i}dx^{j}$ which introduces an additional factor of $\sqrt{-1}$ thus recovering \eqref{eq:calA}.
A helpful feature of this construction is that it also describes the heterotic dual to this local model.
More precisely, it is the linearization obtained from a (singular) $T^{3}$
fibration over $M$.

This alternate presentation already points to an important general point:\ A
priori, there is no reason for the components $\phi_{j}$ in equation
(\ref{complexconnection}) to be simultaneously diagonalizable. Returning to
the PW\ equations, this also means there is no reason to exclude gauge
fluxes through the three-manifold. Let us also note that even if such fluxes
are present, it does not directly mean there will be a bulk four-form flux in
the $G_{2}$ model. This is because these fluxes are inherently localized
on the three-manifold and are \textquotedblleft hard to see\textquotedblright\ from the bulk point
of view. Indeed, the local geometry of the $G_2$ background is primarily sensitive to just
the eigenvalues of $\phi_i$ and bulk $G-$fluxes, and not to any of these non-abelian local features.

A canonical example is the tangent bundle of $T^{\ast}M$. It has the important
feature that the spectral cover description is degenerate.
On the three manifold $M$ we have a vector
bundle with $SU(3)$ structure group and the $\phi$'s certainly do not commute.
Let us note that in heterotic / F-theory duality, the standard embedding also
corresponds to a T-brane configuration of an F-theory compactification
\cite{Aspinwall:1998he,Anderson:2013rka}.

\section{Fluxed PW Solutions \label{sec:BKGND}}

In this section we consider fluxed solutions of the PW system. Our strategy
for obtaining such configurations will be to consider a local description of
the three-manifold as a Riemann surface fibered over an interval, and we shall often
further specialize to the case of a Cartesian product $M = \Sigma \times I$ with $\Sigma$ a Riemann surface
and $I$ an interval. This description will only be valid locally, and so we can either assume these
solutions extend outside of the patch in question, or alternatively, we can
cut off the solution by allowing singular field configurations at
prescribed regions of the three manifold.

To aid in our study of fluxed PW solutions, we shall often assume the
metric on the three-manifold takes the form:%
\begin{equation}
ds_{M}^{2}=g_{tt}dt^{2}+g_{ab}dx^{a}dx^{b},
\end{equation}
where in the above, $t$ denotes a local coordinate on $I$, and $x^{a}$ for
$a=1,2$ denote coordinates on the Riemann surface, and since we often focus on metric independent
questions, we shall also sometimes take the metric to be flat in some local patch.

In terms of this presentation, the PW equations take the form:%
\begin{align}\label{eqn:PWxxx}
F_{ab}-[\phi_{a},\phi_{b}]  &  =0\,,\\
D_{a}\phi_{b}-D_{b}\phi_{a}  &  =0\,,\\
g^{ab}D_{a}\phi_{b}+g^{tt}D_{t}\phi_{t}  &  =0\,,\\
F_{ta}-[\phi_{t},\phi_{a}]  &  =0\,,\\
D_{t}\phi_{a}- D_{a}\phi_{t}  &  =0\,,
\end{align}
modulo unitary gauge transformations. In the above, we have written $D_i = \partial_{i} + A_{i}$ for
the covariant derivative.
We now observe that the first three equations describe a small deformation of
the standard Hitchin system of reference \cite{Hitchin:1986vp}. Indeed, introducing a covariant derivative
$D_{\Sigma}=d+A_{\Sigma}$ and a one-form $\phi_{\Sigma}$ on each fiber,\footnote{More precisely, $D_\Sigma$ and $\phi_\Sigma$ are the pullbacks of $D$ and $\phi$ to the Riemann surface $\Sigma$ that is fibered over the interval $I$.} we
have:%
\begin{align}
F_{\Sigma}-[\phi_{\Sigma},\phi_{\Sigma}]  &  =0\,,\\
D_{\Sigma}\phi_{\Sigma}  &  =0\,,\\
\ast_{\Sigma}\,D_{\Sigma}\ast_{\Sigma}\phi_{\Sigma}  &  =-g^{tt}D_{t}\phi_{t}\,,
\label{deformo}%
\end{align}
which would have described the Hitchin system in the special case where
$g^{tt} D_{t}\phi_{t}=0$.

We remark that these equations tell us that the induced Hitchin system does not describe a Higgs bundle in the mathematician's sense \cite{simpson}; indeed, the condition for harmonicity of the bundle metric is \cite{donaldson, corlette} $D_{\Sigma}\ast_{\Sigma}\phi_{\Sigma}=0$, and equation (\ref{deformo}) tells us that a PW solutions gives a deformation of that condition along the $t$-direction. We point out that the dependence of equation (\ref{deformo}) on the bundle metric is hidden in the definition of the Hodge star $\ast$.

The remaining F-term relations $\mathcal{F}_{ta}=0$ can also be interpreted as a flow
equation:%
\begin{equation}
\mathcal{F}_{ta}=\partial_{t}\mathcal{A}_{a}-\partial_{a}\mathcal{A}%
_{t}+\left[  \mathcal{A}_{t},\mathcal{A}_{a}\right]  =0
\end{equation}
i.e.:%
\begin{equation}
\partial_{t}\mathcal{A}_{a}+\left[  \mathcal{A}_{t},\mathcal{A}_{a}\right]
=\partial_{a}\mathcal{A}_{t}.
\end{equation}

Geometrically, we interpret the flow equations as a gluing construction for
local Calabi--Yau threefolds. To see why, we first recall the correspondence
between the Hitchin system on a genus $g$ curve $\Sigma$ and the integrable system associated to a family of non-compact Calabi--Yau manifolds each containing $\Sigma$ as a curve of
ADE\ singularities \cite{Diaconescu:2006ry,Diaconescu:2005jw,Anderson:2017rpr}. Let $Y$ denote the Calabi--Yau threefold that is the central fiber of the deformation family. Recall that the isomorphism between the integrable systems implies in particular that variations in the complex structure for $Y$ are described by (spectral curves of) Higgs fields $\varphi$ - i.e., adjoint-valued $(1,0)$-forms on the curve $\Sigma$.
With notation as in equation (\ref{rhoC}) for harmonic $(1,1)$ forms on the
ADE\ singularity, the variations of the holomorphic three-form on the local Calabi--Yau decompose as:
\begin{equation}
\delta\Omega_{(2,1)}=\varphi_{\alpha}\wedge\omega^{\alpha}\text{.}%
\end{equation}
The Calabi--Yau condition enforces the condition $\overline{\partial}(\delta\Omega_{(2,1)})=0$,
which translates to one of the Hitchin equations:%
\begin{equation}
\overline{\partial}_{A}\varphi=0\text{.}%
\end{equation}
The deformation of line (\ref{deformo}) tells us that the righthand side is no
longer zero. Translating back to the Calabi--Yau integrable system, we see that
we instead get a deformation of a Calabi--Yau manifold that does not respect the K\"ahler condition, resulting in a ``symplectic Calabi--Yau'' in the sense of Smith, Thomas and Yau \cite{thomas} (for a recent discussion see reference \cite{Andriolo:2018yrz}, and for a review see reference \cite{Grana:2005jc} and additional references therein).

From the perspective of the local $G_{2}$, we thus see that asymptotically
near the boundaries of the interval, we retain an approximate Calabi--Yau
geometry, but as we proceed to the interior of the interval, each fiber will
instead be described by a symplectic Calabi--Yau.

Let us illustrate the correspondence between the spectral equation for
the Higgs field and deformations of its dual local Calabi--Yau in the case
of $G=SU(N)$. The spectral equation in the fundamental representation is:
\begin{equation}
\det(u_{\text{hol}}\mathbb{I}_{N} - \varphi_{\text{hol}} ) = 0,
\end{equation}
with $u_{\text{hol}}$ a local coordinate in the cotangent direction of $T^{\ast} \Sigma$. Since we also have a Higgs field in the more general
case, it is natural to consider two related spectral equations, as generated by the PW system. First, we have the one closely linked to the
asymptotic Hitchin system associated with $\Phi = i \phi$:
\begin{equation}\label{specbound}
\det(u_{z}\mathbb{I}_{N} - \Phi_{z} ) = 0.
\end{equation}
This equation only makes sense asymptotically, since in the bulk of the three-manifold, the Higgs field of the PW system is not a
holomorphic (or even meromorphic) section of a bundle on the curve $\Sigma$.
Indeed, more generally we ought to speak of the spectral equations:
\begin{equation}\label{specbulk}
\det(u_{i}\mathbb{I}_{N} - \Phi_{i} ) = 0.
\end{equation}
where here, $u_{i}$ for $i=1,2,3$ are coordinates in the cotangent direction of $T^{\ast} M$.
This is a triplet of real equations in $T^{\ast} M$ which cut out a three-manifold in this ambient space.
Geometrically, then, we can interpret the spectral equation of line (\ref{specbulk}) as a special Lagrangian manifold
in $T^{\ast}M$ with boundaries specified by the holomorphic curves dictated by equation (\ref{specbound}). Indeed,
the PW equations ensure that supersymmetry has been preserved.

This also allows us to elaborate on the sense in which having $[\Phi_i , \Phi_j] \neq 0$ is an example of T-brane phenomena. Even though each
$\Phi$ takes values in the unitary Lie algebra $\mathfrak{g}$, it is also natural to consider linear combinations
which take values in the complexification $\mathfrak{g}_{\mathbb{C}}$. If the complexified combination $\Phi_z = \Phi_1 + i \Phi_2$ is a nilpotent element of $\mathfrak{g}_{\mathbb{C}}$, then we observe that the holomorphic spectral equation of line (\ref{specbound}) is degenerate.
Indeed, even though $\Phi_1$ and $\Phi_{2}$ are hermitian (and thus never nilpotent), their complex combination can of course be nilpotent. From the perspective of a global $G_2$ background, this also suggests that such phenomena may be invisible to the geometry, instead being encoded in non-abelian degrees of freedom localized along lower-dimensional subspaces. A related
comment is that even though each component of the Higgs field $\Phi$ is hermitian, when these matrices
do not commute, there is no canonical way in which we can speak of the spectral sheets intersecting along a locus of
symmetry enhancement. We return to these issues in section \ref{sec:ALGEBRAIC}.

The plan in the remainder of this section is as follows.
First, we discuss some general aspects of background
solutions in a local patch.  Starting from a solution
along a single fiber, we show that this solution extends to a local
neighborhood of the three-manifold. We then present some explicit
examples of backgrounds, including T-brane configurations.

\subsection{Background Solutions in a Local Patch}

We first locally characterize solutions of the system in a patch
with trivial topology. By abuse of notation, we continue to write $M$ for this
local patch. The F-term equations of motion tell us that we are dealing with a
complexified flat connection, so the most general solution for the gauge
connection is of the form:%
\begin{equation}
\mathcal{A}=g^{-1}dg,
\end{equation}
where $g:M\rightarrow G_{\mathbb{C}}$ takes values in the complexified gauge
group. One must remember that here, we are working with a complexified
connection, so even though the gauge field appears to be \textquotedblleft
pure gauge,\textquotedblright\ we must only permit gauge transformations which
do not alter the asymptotic behavior of the gauge field. A related point is
that to actually find a solution in unitary gauge, we need to substitute
our expression for $\mathcal{A} = g^{-1} dg $ into the D-term constraint $g^{ij}D_{i} \phi_{j} = 0$,
resulting in a second order partial differential equation for $g(x)$.

As an illustrative example, suppose that $g$ is of the form:%
\begin{equation}
W \cdot g = 1+h+\frac{h^{2}}{2!}+...,
\end{equation}
where $W$ is a general element of $G_{\mathbb{C}}$, and we have introduced an
infinitesimal $h:M\rightarrow\mathfrak{g}_{\mathbb{C}}$ a Lie algebra valued
function on the patch. Then, the complexified connection takes the form:%
\begin{equation}
\mathcal{A}=dh+...\text{.}%
\end{equation}
Feeding this into the D-term constraint, we get, to leading order in $h$:%
\begin{equation}
g^{ij}\mathcal{D}_{ij}=g^{ij}\partial_{i}\partial_{j}(h-h^{\dag})=0.
\end{equation}
So in other words, $\operatorname{Im}h$ is a Lie algebra valued harmonic map.

There are, of course, more general choices for $g(x)$ and we will in fact find it necessary to
consider a more general choice of complexified connection to generate novel examples of T-brane
configurations with localized matter.

\subsection{Power Series Solutions \label{ssec:powseries}}

We now turn to a more systematic method for building up solutions of the PW equations in the presence of flux.
The basic idea is that we shall split up our three-manifold as a local product $\Sigma \times I$ with
$\Sigma$ a Riemann surface (possibly with punctures) and $I$ an interval. We introduce a local coordinate on
$I$ denoted by $t$ so that $t = 0$ is in the interior of the interval, and coordinates $x_{a}$ for $a =1,2$ for coordinates
on the Riemann surface. It will sometimes prove convenient to also use coordinates $z = x_1 + ix_2 = x + iy$.

We show that if enough initial data is specified
at $t = 0$, then we can start to extend this solution in a neighborhood, building up a solution on the entire patch of the three-manifold.
The demonstration of this will be to develop a power series expansion in the variable $t$ for all fields of the PW system,
and solve the expanded equations order by order in this parameter. We shall not concern ourselves with whether
the series converges, because we do actually anticipate that there could be singular behavior for the fields
at locations on the three-manifold. This is additional physical data, and must be allowed to make sense of the most general
configurations of relevance for physics. As a final comment, while we cannot exclude the possibility of solutions which do not
admit a power series expansion in some local neighborhood, we expect on physical grounds that such solutions are likely pathological as they would significantly a breakdown of the $G_2$ structure at more than just a higher codimension subspace.

We consider a series expansion of the complexified connection around $t = 0$,
\begin{equation}\label{eqn:PSexp}
  \mathcal{A}_i(t, x_a) = \sum_{j=0}^\infty \mathcal{A}_i^{(j)}(x_a) t^j \,.
\end{equation}
Furthermore, it is convenient to work in ``temporal gauge'' with:
\begin{equation}
  A_t^{(j)} = 0 \,.
\end{equation}
In this gauge, and with the power series expansion (\ref{eqn:PSexp}), the
PW equations can be written as non-trivial differential
equations on the coefficients
\begin{equation}\label{eqn:PSnoteasy}
  \begin{gathered}
    \sum_{j = 0}^\infty \left( \partial_a A_b^{(j)} - \partial_b A_a^{(j)} +
    \sum_{m=0}^j \left( [ A_a^{(j-m)}, A_b^{(m)}] - [\phi_a^{(j-m)},
    \phi_b^{(m)}] \right) \right) t^j = 0 \cr
    \sum_{j = 0}^\infty \left( \partial_a \phi_b^{(j)} + \sum_{m=0}^j
    [A_a^{(j-m)}, \phi_b^{(m)}] - \partial_b \phi_a^{(j)} - \sum_{m=0}^j
    [A_b^{(j-m)}, \phi_a^{(m)}] \right) t^j = 0 \,,
  \end{gathered}
\end{equation}
together with equations which fix the higher order coefficients in terms of
the preceding ones,
\begin{equation}\label{eqn:PSeasy}
  \begin{gathered}
    \sum_{j = 0}^\infty \left( g^{ab} \left( \partial_a\phi_b^{(j)} +
    \sum_{m=0}^j [ A_a^{(j-m)}, \phi_b^{(m)} ] \right) +
    (j+1)g^{tt}\phi_t^{(j+1)} \right) t^j = 0 \cr
    \sum_{j = 0}^\infty \left( (j+1)A_a^{(j+1)} - \sum_{m=0}^j [
    \phi_t^{(j-m)}, \phi_a^{(m)} ] \right) t^j = 0 \cr
    \sum_{j = 0}^\infty \left( (j+1)\phi_a^{(j+1)} - \partial_a \phi_t^{(j)} -
    \sum_{m=0}^j [ A_a^{(j-m)}, \phi_t^{(m)} ] \right) t^j = 0 \,.
  \end{gathered}
\end{equation}
At $t=0$ these equations collapse to
\begin{equation}
  \begin{gathered}
    F_{ab}^{(0)} - [\phi_a^{(0)}, \phi_b^{(0)}] = 0 \cr
    D_a^{(0)} \phi_b^{(0)} - D_b^{(0)} \phi_a^{(0)} = 0 \cr
    g^{ab}D_a^{(0)}\phi_b^{(0)} + g^{tt}\phi_t^{(1)} = 0 \cr
    A_a^{(1)} - [\phi_t^{(0)}, \phi_a^{(0)}] = 0 \cr
    \phi_a^{(1)} - D_a^{(0)} \phi_t^{(0)} = 0 \,.
  \end{gathered}
\end{equation}
We will assume that $A_a^{(0)}$ and $\phi_a^{(0)}$ are such that the
non-trivial zeroth order differential equations are solved, and the higher
order coefficients are fixed by solving the linear equations. The one
remaining free parameter is $\phi_t^{(0)}$, which sets the ``trajectory'' of
the solution. Once we are given this initial set of data, solving the zeroth
order equations, we can show that the PW equations are solved to
all orders in $t$.

To show that solving the zeroth order equations leads to a solution at all
orders in the power series expansion we will first substitute
(\ref{eqn:PSeasy}) into (\ref{eqn:PSnoteasy}). We begin by
noticing that the commutators that appear in the differential equations
(\ref{eqn:PSnoteasy}), for the $(j+1)$-term, can always be written as
\begin{equation}
  \sum_{m=0}^{j+1} [\lambda_a^{(j+1-m)}, \chi_b^{(m)}] =
  \frac{1}{j+1}\sum_{m=0}^j \left( (j+1-m)[\lambda_a^{(j+1-m)},
  \chi_b^{(m)}] + (m+1)[\lambda_a^{(j-m)}, \chi_b^{(m+1)}] \right) \,,
\end{equation}
where $\lambda$ and $\chi$ represent either $A$ or $\phi$. We will then
replace, using (\ref{eqn:PSeasy}), the terms $\lambda_a^{(j+1-m)}$ and
$\chi_b^{(m+1)}$. Combining this expansion with some double sum identities,
together with the Jacobi identity, one finds, after some algebra,
\begin{multline}\label{eqn:PSd1}
  \partial_a A_b^{(j+1)} - \partial_b A_a^{(j+1)} + \sum_{m=0}^{j+1} \left( [
  A_a^{(j+1-m)}, A_b^{(m)}] - [\phi_a^{(j+1-m)}, \phi_b^{(m)}] \right) \cr
  = \frac{1}{j+1} \sum_{k=0}^{j} \left[ \phi_t^{(k)}, \partial_a
    \phi_b^{(j-k)} - \partial_b \phi_a^{(j-k)} + \sum_{m=0}^k \left(
    [A_a^{(j-k-m)}, \phi_b^{(m)}] - [ A_b^{(j - k - m)}, \phi_a^{(m)}] \right)
  \right] \,,
\end{multline}
and
\begin{multline}\label{eqn:PSd2}
  \partial_a \phi_b^{(j+1)} + \sum_{m=0}^{j+1}
  [A_a^{(j+1-m)}, \phi_b^{(m)}] - \partial_b \phi_a^{(j+1)} - \sum_{m=0}^j
  [A_b^{(j+1-m)}, \phi_a^{(m)}] \cr
  = \frac{1}{j+1} \sum_{k=0}^{j} \left[ \phi_t^{(k)},
  \partial_a A_b^{(j-k)} - \partial_b A_a^{(j-k)} +
  \sum_{m=0}^{j-k} \left( [ A_a^{(j-k-m)}, A_b^{(m)}] - [\phi_a^{(j-k-m)},
    \phi_b^{(m)}] \right)
  \right] \,.
\end{multline}
If we define the shorthand notation where the power series expansion in
(\ref{eqn:PSnoteasy}) look, respectively, like
\begin{equation}
  \sum_{j=0}^\infty G_{ab}^{(j)} t^{(j)} \,, \quad \sum_{j=0}^\infty
  H_{ab}^{(j)} t^j \,,
\end{equation}
then we can immediately see, from (\ref{eqn:PSd1}) and (\ref{eqn:PSd2}), that,
after plugging in the solutions to the linear equations (\ref{eqn:PSeasy}),
\begin{equation}
  \begin{aligned}
    G_{ab}^{(j+1)} &= \frac{1}{j+1} \sum_{k=0}^j [ \phi_t^{(k)},
    H_{ab}^{(j-k)} ] \cr
    H_{ab}^{(j+1)} &= - \frac{1}{j+1} \sum_{k=0}^j [ \phi_t^{(k)},
    G_{ab}^{(j-k)}] \,.
  \end{aligned}
\end{equation}
These expressions make obvious the inductive proof that if
\begin{equation}
  G^{(0)}_{ab} = 0 \,, \quad H^{(0)}_{ab} = 0 \,,
\end{equation}
which we assume, then it follows that
\begin{equation}
  G^{(j)}_{ab} = 0 \,, \quad H^{(j)}_{ab} = 0 \,,
\end{equation}
Thus, if we have given $A_a^{(0)}$ and $\phi_a^{(0)}$ such that the zeroth order
equations in (\ref{eqn:PSnoteasy}) are solved, then we can construct a full
solution of the PW equations by specifying all the higher order
coefficients as in (\ref{eqn:PSeasy}). Note that $\phi_t^{(0)}$ is
unspecified, and we consider this parameter as determining how
the solution extends to a solution of the full system. Further note
that we did not require the first equation in (\ref{eqn:PSeasy}) to arrive at
the conclusions (\ref{eqn:PSd1}) and (\ref{eqn:PSd2}).

Resumming this power series can be viewed as a complexified gauge transformation. To see why, consider a (partial) solution of the Hitchin system at $t=0$ and generate a flow along the interval $I$ to produce a non-trivial dependence in the $t$ direction. This has the advantage of automatically solving the F-term equations of motion and gives a method for iteratively solving the D-term equations of motion. We start with a complexified connection $\mathcal A^{(0)}( x_1,x_2)$ with legs only on $\Sigma$ which gives a solution to the F-term condition $\mathcal F_{12} = 0$. Note that we do not require it to be a full solution to the PW equations and in particular we shall be interested in the case when $\mathcal A^{(0)}(x_1,x_2)$ does not give vanishing D-terms. This issue can be addressed by considering a complexified gauge transformation by a gauge parameter $\exp \chi$ of the form\footnote{To avoid changing the solution at $t=0$ we will take $\chi_0(x_1,x_2) = 0$. Moreover we will assume that the Lie-algebra valued functions $\chi_i(x_1,x_2)$ are suitably chosen to ensure that the gauge choice $A_t =0$ is enforced.}
\begin{align}
\exp \chi = \exp \left[\sum_i \chi_i (x_1,x_2) t^{i} \right]\,.
\end{align}
After performing this complexified gauge transformation the background will still solve the F-term equations. Moreover it also admits a power series expansion around $t=0$, and borrowing from the results of section \ref{ssec:powseries}, it is possible to find a solution of the D-term equations by suitably choosing the terms $h_i(x_1,x_2)$ in the complexified gauge transformation.

\subsection{Examples of Backgrounds}

In this section we present some examples of background solutions.
We first discuss some abelian examples in which no gauge field
flux is switched on, and then present a novel non-abelian
solution with flux.

\subsubsection{First Abelian Background \label{ssec:ABEL}}

\label{sec:abackground}
Our first example is a particular case of the solutions already constructed in
\cite{Pantev:2009de}. We take an $SU(2)$ gauge theory and write the
complexified connection as
\begin{equation}
\mathcal{A}=\left[
\begin{array}
[c]{cc}%
\alpha & \\
& -\alpha
\end{array}
\right]  \,,
\end{equation}
where:%
\begin{equation}
\alpha=zd\overline{z}+\overline{z}dz-4t dt=2xdx+2ydy-4tdt,
\end{equation}
with $z=x+iy$. This background solves the F-term equations by virtue of the
fact that $\mathcal{A}=g^{-1}dg$ with
\[
g=\exp\left(
\begin{array}
[c]{cc}%
z\bar{z}-2t^{2} & 0\\
0 & 2t^{2}-z\bar{z}%
\end{array}
\right)  \,.
\]
The D-term equations are satisfied as well because the function $z\bar
{z}-2t^{2}$ is harmonic in $\mathbb{C}\times\mathbb{R}_{t}$ with a flat metric.

In terms of the fields $A$ and $\Phi = i\phi$, the gauge field $A=0$, and the Higgs
field is given by:\footnote{To extract $\Phi$  and $A$ from the complexified connection $\mathcal A$ in a unitary gauge it suffices to take its hermitian and anti-hermitian parts respectively.}
\begin{equation}
\Phi=\left[
\begin{array}
[c]{cc}%
df & \\
& -df
\end{array}
\right]
\end{equation}
with:
\begin{equation}
f=z\overline{z}-2t^{2}.
\end{equation}
We note that the Hessian of $f$ has signature $(+,+,-)$ in the $(x,y,t)$
coordinate system, so it translates to a geometry with a codimension seven
singularity in the local $G_{2}$ background.

\subsubsection{Second Abelian Background
\label{ssec:CODIM6}}

As another example, we can also take the abelian background:%
\begin{equation}
\mathcal{A}=\left[
\begin{array}
[c]{cc}%
\beta & \\
& -\beta
\end{array}
\right]  \,,
\end{equation}
with:%
\begin{equation}
\beta=zdz+\overline{z}d\overline{z}=2xdx-2ydy.
\end{equation}
The corresponding Higgs field in this case is:%
\begin{equation}
\Phi=\left[
\begin{array}
[c]{cc}%
df & \\
& -df
\end{array}
\right]
\end{equation}
with:
\begin{equation}
f=\frac{1}{2}z^{2}+\frac{1}{2}\overline{z}^{2},
\end{equation}
We note that the Hessian of $f$ has signature $(+,-,0)$ in the $(x,y,t)$
coordinate system, so it translates to a geometry with a codimension six
singularity in the local $G_{2}$ background. Starting from this example, we
obtain a codimension seven singularity by adding a perturbation
$f_{\text{pert}}$ which also has vanishing Hessian:%
\begin{equation}
f_{\text{new}}=f + f_{\text{pert}}.
\end{equation}
We can again solve the F- and D-terms for this system, and thus obtain a
genuine background in this more general case as well.

\subsubsection{Fibering a Hitchin System \label{ssec:HITPERT}}

Building on this previous example, we can also consider more general
backgrounds on $\Sigma\times I$ by first solving the Hitchin system on a curve
$\Sigma$, and then adding perturbations so that it has a non-trivial profile
on the interval. In practice, we accomplish this by starting with a
complexified connection $\mathcal{A}_{\text{H}}$ defined on $\Sigma$ which
solves the Hitchin equations, and trivially extending this to $\Sigma\times
I$. This solves our fluxed PW\ equations in the degenerate limit where
$g^{tt}\rightarrow0$, corresponding to the physical limit where we take the
curve $\Sigma$ much smaller than the interval. Adding a perturbation
$\mathcal{A}_{\text{pert}}$ to the complexified connection:%
\begin{equation}
\mathcal{A}_{\text{new}}=\mathcal{A}_{\text{H}} + \mathcal{A}%
_{\text{pert}},
\end{equation}
we observe that the power series analysis of section \ref{ssec:powseries} ensures that we can
consistently add in such perturbations and produce a solution to the full
PW\ system of equations. Indeed, the main freedom we have in specifying this
contribution is the $t$-dependence which is wholly absent from $\mathcal{A}%
_{\text{H}}$.

\subsubsection{Non-Abelian Background \label{sec:nabackground}}

We can also consider the limit where the deformation away from a Hitchin system is large.
This will be our main example of a T-brane configuration. In this solution we take
an $SU(3)$ gauge theory though the technique employed here can be easily
generalized to other non-abelian gauge groups with rank greater than one. Our analysis
follows a similar treatment to that presented in reference \cite{Cecotti:2010bp}.
When written in the fundamental representation the background fields are:
\begin{align}
\phi &  =\left[
\begin{array}
[c]{ccc}%
\frac{i}{3}dh & -\upsilon\,\bar{z}e^{-f(z,\bar{z})}d\bar{z}+\varepsilon
\,e^{f(z,\bar{z})}dz & 0\\
\upsilon ze^{-f(z,\bar{z})}dz-\varepsilon\,e^{f(z,\bar{z})}d\bar{z} & \frac
{i}{3}dh & 0\\
0 & 0 & -\frac{2i}{3}dh
\end{array}
\right]  \,,\\
A  &  =\left[
\begin{array}
[c]{ccc}%
a & 0 & 0\\
0 & -a & 0\\
0 & 0 & 0
\end{array}
\right], \,\,\,\text{where}\,\,\, a=\frac{i}{2}\left(  \partial_{\bar{z}}f(z,\bar{z})d\bar{z}-\partial
_{z}f(z,\bar{z})dz\right)  .
\end{align}

We will shortly show that for suitable choices of the functions $h(z,\overline
{z},t)$ and $f(z,\overline{{z}},t)$, this background indeed solves the F- and
D-term equations of the PW system. Our main interest will be in picking background
values so that as much as possible can be ``hidden'' from the classical geometry.
In particular, if we take $h$ to be a function with a degenerate Hessian (one
zero eigenvalue), a PW solution with just this
contribution would appear to support a
codimension six singularity along the locus $dh=0$. Adding in additional
off-diagonal components to the Higgs field need not change this interpretation. For
example, if we consider the purely off-diagonal contributions to the
$z$ component of the Higgs field, namely $\Phi^{\text{off-diag}}_{z}$, we
observe that when $\varepsilon \rightarrow 0$, the limiting Hitchin system has a
degenerate spectral equation, i.e., there is not even a codimension six contribution to matter localization
from these holomorphic off-diagonal contributions. Taken together, this suggests that the proper geometric
interpretation will appear to contain at most a codimension six singularity.

Let us now turn to an explicit example. The background will satisfy the
supersymmetry conditions provided that the two functions $h(z,\bar z,t)$ and
$f(z,\bar z)$ satisfy the differential equations:
\begin{align}\label{eq:PIII}4\partial_{z} \partial_{\bar z} h + \partial_{t}^{2} h = 0\,,\\
\partial_{z} \partial_{\bar z} f = \varepsilon^{2} e^{2f}- \upsilon^{2}
|z|^{2} e^{-2f}\,.\label{eq:PIIIprime}
\end{align}
The first equation allows for several solutions, and in the following we shall
take:
\begin{align}
h = \frac{\kappa}{8} (z+\bar z)^{2}-\frac{\kappa}{2}t^{2}\,.
\end{align}
The second equation can be related to a Painlev\'e III transcendent via
suitable change of variables provided that $\varepsilon\neq0$. In this case,
the solution has an asymptotic expansion near $z=0$ of the form:
\begin{align}
f(r) = \log c+\frac{1}{3} \log\upsilon-\frac{2}{3} \log\varepsilon+ c^{2}
r^{2} + \mathcal{O}(r^{4})\,,
\end{align}
where $r^{2} = \varepsilon^{\frac{2}{3}} \upsilon^{\frac{2}{3}}z \bar z$ and
$c$ is a real constant. In order to avoid singularities at finite values of
$r$ one should fix:
\begin{align}
c = 3^{1/3} \frac{\Gamma\left[  \frac{2}{3}\right]  }{\Gamma\left[  \frac
{1}{3}\right]  } \sim0.73\,.
\end{align}
In the following we shall be also interested in the case where $\varepsilon=0$
because in this case some of the effect of the Higgs field background would
become nilpotent. When this happens \eqref{eq:PIIIprime} becomes a modified
Liouville equation with solution:
\begin{align}
f(z,\bar z) = - \log\left[  \frac{2 }{1-\upsilon^{2}|z|^{4}}\right]  \,.
\end{align}

\subsubsection{More General Embeddings \label{ssec:more}}

As a final comment, we can also generalize the example of section
\ref{sec:nabackground} to other choices of gauge groups. Observe that the
Higgs field of the previous example takes values in an $\mathfrak{su}(2)
\times\mathfrak{u}(1)$ subalgebra of $\mathfrak{su}(3)$. More generally, we
can specify a homomorphism $\mathfrak{su}(2) \rightarrow\mathfrak{g}$, and a
commuting $\mathfrak{u}(1)$ subalgebra. Then, we clearly also generate a
broader class of examples of fluxed PW solutions.

\section{Examples of Localized Matter \label{sec:MATTER}}

In the previous section we discussed a general method for generating
consistent solutions to the PW\ equations, and its connection to TCS-like
constructions of $G_{2}$ manifolds. Assuming the existence of a consistent
background, we would now like to determine whether localized zero modes are present.

To frame the discussion to follow, we assume that the complexified connection
takes values in some maximal subalgebra $\mathfrak{h}^{\prime}_{\mathbb{C}}\subset\mathfrak{g}%
_{\mathbb{C}}$ so that the commutant subalgebra is $\mathfrak{h}_{\mathbb{C}}%
\subset\mathfrak{g}_{\mathbb{C}}$. We specify a zero mode by considering
fluctuations around this background. The presence of $\mathcal{N}=1$ chiral
multiplets can be understood in terms of fluctuations of the complexified
connection $\mathcal{A}$. We therefore consider the expansion:%
\begin{equation}
\mathcal{A=A}^{(0)}+\psi.
\end{equation}
In what follows, we shall omit the superscript $(0)$ to avoid overloading the
notation. Note that under an infinitesimal gauge transformation, the zero mode
solution will shift as:%
\begin{equation}
\psi\rightarrow\psi+D_{\mathcal{A}}\chi, \label{gaugetrans}%
\end{equation}
where $\chi$ is an adjoint valued zero-form in the Lie algebra.

We need to study the various representations appearing in the unbroken gauge
algebra. This is by now a standard story which is largely borrowed from the
case of compactifications of the heterotic string as well as local F-theory
models so we shall be brief. Decomposing the adjoint representation of
$\mathfrak{g}_{\mathbb{C}}$ into irreducible representations of $\mathfrak{h}_{\mathbb{C}%
}\times \mathfrak{h}^{\prime}_{\mathbb{C}}$, we have:%
\begin{align}
\mathfrak{g}_{\mathbb{C}}  &  \supset \mathfrak{h}_{\mathbb{C}}\times \mathfrak{h}^{\prime}_{\mathbb{C}}\\
\text{ad}\left(  \mathfrak{g}_{\mathbb{C}}\right)   &  =\underset{i}{%
{\displaystyle\bigoplus}
}\left(  \mathcal{T}_{i},\mathcal{R}_{i}\right)  . \label{irreps}%
\end{align}
For a prescribed representation $\mathcal{R}$ of $\mathfrak{h}^{\prime}_{\mathbb{C}}$, we
therefore need to consider the action of the complexified connection
$\mathcal{A}_{\mathcal{R}}$ on the zero mode.

There are various ways to analyze the zero mode content of this theory. Most
directly, we can return to the PW equations, and expanding around
a given background we can seek out zero modes modulo unitary gauge
transformations. This approach makes direct reference to the metric on the
three manifold and is certainly necessary if we want the explicit wave
function profile for the zero modes. The relevant linearization of the
PW\ system of equations is:
\begin{align}
\partial_{i}{\psi}_{j}-\partial_{j}{\psi}_{i}+\left[  {\psi}_{i}%
,\mathcal{A}_{j}\right]  +\left[  \mathcal{A}_{i},{\psi}_{j}\right]   &  =0\\
g^{ij}\left(  \partial_{i}\overline{{\psi}}_{j}-\partial_{j}{\psi}_{i}+\left[
{\psi}_{i},\overline{\mathcal{A}}_{j}\right]  +\left[  \mathcal{A}%
_{i},\overline{{\psi}}_{j}\right]  \right)   &  =0
\end{align}
modulo unitary gauge transformations as in line (\ref{gaugetrans}). In
practice, we shall actually demand that our zero modes satisfy a slightly
stronger set of conditions:
\begin{align}
\partial_{i}{\psi}_{j}-\partial_{j}{\psi}_{i}+\left[  {\psi}_{i}%
,\mathcal{A}_{j}\right]  +\left[  \mathcal{A}_{i},{\psi}_{j}\right]   &  =0\\
g^{ij}\left(  \partial_{i}{\psi}_{j}+\left[  \overline{\mathcal{A}}_{i} ,
{\psi}_{j} \right]  \right)   &  =0.
\end{align}
Any solution to this set of equations automatically produces a solution to the
linearized PW equations. It is convenient to use these zero mode equations to
track their falloff and thus to ensure they are actually normalizable.
This, for example, is what allows us to determine whether we have a
normalizable mode in a representation $\mathcal{R}$ or the complex conjugate
representation $\mathcal{R}^{c}$.

Now, one unpleasant feature of this approach is that it often requires dealing
with coupled partial differential equations. In the special case where
$\mathcal{A}_{\mathcal{R}}$ is diagonal, this is not much of an issue, but in
more general T-brane backgrounds this can lead to significant technical complications.

At a qualitative level, however, it is straightforward to see how to pick
appropriate backgrounds which could generate localized matter. First of all,
we can attempt to find a localized zero mode in the Hitchin system. In
M-theory language, this would produce a 5D hypermultiplet. These multiplets
can all be organized according to 4D $\mathcal{N} = 2$ hypermultiplets. In the
presence of a non-trivial field profile on the transverse direction to the 4D
spacetime, this leads to a localized chiral mode, producing a single 4D
$\mathcal{N} = 1$ chiral multiplet. We see how this comes about in our zero
mode equations by taking $g^{tt}$ small relative to the other factors of the
metric on $\Sigma\times I$. Indeed, in this case, the leading order behavior
is governed by a mild deformation of the zero mode equations on the curve
$\Sigma$, and then there is a broadly localized mode in the remaining
$t$-direction. That being said, it is clear that there is some level of
``backreaction'' in the form of these zero mode solutions, so obtaining
explicit wave functions in this approach is more challenging.

Our plan in the remainder of this section will be to discuss some general
features of zero mode solutions in T-brane backgrounds, and to then discuss
explicit examples. In section \ref{sec:ALGEBRAIC} we give a conjectural algebraic
method of analysis for detecting the presence of localized zero modes.

\subsection{Cohomological Approach}\label{sec:cohom}

One approach that may be employed in the search for a solution of the wave function equations involves relying on the cohomology\footnote{Here and in what follows, we will always consider $L^2$-cohomology.} of the operator $D_{\mathcal A}$. Let $E$ denote a complex vector bundle with connection $\mathcal A$, and $E_{\mathcal R}$ the bundle associated to a representation $\mathcal R$ of the gauge group $G_\mathbb{C}$. By virtue of the vanishing of the F-term conditions the operator $D_{\mathcal A} : \Omega^{\bullet}(M,E_{\mathcal R}) \rightarrow \Omega^{\bullet+1}(M,E_{\mathcal R}) $ squares to zero implying that we can consider its cohomology complex, much as in
reference \cite{Braun:2018vhk}.

The linearized F-terms modulo complex gauge transformations are solved by a cohomology class $c \in H^1_{ D_{\mathcal A}} (M)$, and the linearized D-term constraint requires us to find a representative for $c$ that is annihilated by the dual operator $D^{\dag}_{\mathcal A}\equiv \ast\,  D_{\overline {\mathcal A}} \,\ast :\Omega^{\bullet}(M,E_{\mathcal R}) \rightarrow \Omega^{\bullet-1}(M,E_{\mathcal R})$. For a \emph{closed} $3$-manifold, a standard integration by parts argument says that the solution is given by the harmonic representative for $c$ (we remind the reader that given any elliptic complex $\mathcal{E}^\bullet$ with metric over a compact manifold, there is a Hodge isomorphism $\mathcal{H}^k(\mathcal{E}^\bullet) \cong H^k(\mathcal{E}^\bullet)$.

The situation is trickier when $M$ is allowed to be non-compact: not only does one have to choose appropriate decay conditions on the fields in order for the boundary term to vanish, but also one must work with metrics such that the relevant cohomology classes admit harmonic representative(s). In fact, even when one considers the simplest elliptic operator - the Hodge Laplacian $\Delta = dd^*+d^*d$ - on a non-compact Riemannian manifold, the space of $L^2$-harmonic forms $\mathcal{H}^k_\Delta (M)$ depends strongly on the metric. For example, $\mathcal{H}^0_\Delta (M)$ is either $\mathbb{R}$ or $\left\{ 0 \right\}$ depending on whether $\text{vol}(M) < \infty$ or $\text{vol}(M)= \infty$. However, one can give a cohomological interpretation for $\mathcal{H}^k_\Delta (M)$ for special metrics; in particular, Atiyah, Patodi and Singer \cite{Atiyah:1975jf} proved that for manifolds with \emph{cylindrical ends}\footnote{A $n$-dimensional Riemannian manifold $M$ has cylindrical ends if there is a $n$-dimensional compact submanifold $K \subset M$ with smooth non-empty boundary $\partial K$, such that $M\setminus K$ is isometric to $\partial{K} \times (0,\infty)$.}, $\mathcal{H}^k_\Delta(M) \cong \text{Im}(H^k_c(M) \to H^k(M))$, i.e. the image of compactly supported cohomology in absolute cohomology.

In our present situation with $M = \Sigma \times I$, we have, under mild assumptions, an example of a cylindrical manifold. Assuming our solution $\psi$ to the linearized F-terms vanishes at infinity $M_\infty \subset \overline{M}$, we can package our solution as a relative cohomology class $c \in H^1_{D_{\mathcal A}}(M,M_\infty)$. Suppose that the order of vanishing \emph{and} the metric are chosen compatibly so as to cancel the boundary term, and moreover that the result of Atiyah, Patodi and Singer generalizes to $D_{\mathcal A}$-harmonic one-forms. Then, a solution to the PW equations on $M$ is given by a $D_{\mathcal A}$-harmonic representative for $c$.

This suggests a practical way to try and find a solution of the wave function equations of motion: starting with any solution $\psi_0$ to the F-term wave function equations we handily build other solutions $\psi_\chi $ as the gauge transformed $\psi_0$ via an element $\chi \in \Omega^0 (M,E_{\mathcal R})$, that is $\psi_\chi = \psi_0 - D_{\mathcal A} \chi$. Thus, even if $\psi_0$ fails to solve the linearized D-term equation it is possible to find a suitable gauge transformed $\psi_\chi$ that is a solution of the full system, which allows us to recast the D-term condition as an equation on $\chi$
\begin{align}
\label{eq:Dtchi}
D^\dag_{ {\mathcal A}}  \psi_\chi = 0 \, \quad \implies \quad \Delta_{\mathcal A} \chi = D_{\mathcal A}^\dag \psi_0 \,,
\end{align}
where we defined the Laplacian $\Delta_{\mathcal A} \equiv D^{\dag}_{\mathcal A} D_{\mathcal A} + D_{\mathcal A} D^{\dag}_{\mathcal A}$. While we are not in general able to show that a suitable $\chi$ solving \eqref{eq:Dtchi} exists we can push this analysis further and argue that when a solution to \eqref{eq:Dtchi} exists, then it is sufficient to check that the mode $\psi_0$ is normalizable to establish existence of a zero mode. This is due to the fact that the harmonic representative in any cohomology class will minimize the norm. Indeed, calling $\psi_{\text{hrm}}$ the harmonic representative we find that the norm of any other element in the cohomology class $\psi = \psi_{\text{hrm}} + D_{\mathcal A} \chi$ is
\begin{align}
\int_{M} || \psi||^2 = \int_{M} || \psi_{\text{hrm}} + D_{\mathcal A} \chi||^2 =  \int_{M} || \psi_{\text{hrm}}||^2 +|| D_{\mathcal A} \chi||^2 > \int_{M} || \psi_{\text{hrm}}||^2\,,
\end{align}
so in other words, the (already normalizable) trial wave function has bigger norm than the harmonic representative.

\subsection{First Abelian Example \label{ssec:ABELIAN}}

We first take a look at the zero modes in the background introduced in Section
\ref{sec:abackground}. In this example we took the gauge group to be $SU(2)$
and its adjoint representation decomposes as:%
\begin{align}
\mathfrak{su}(2)  &  \supset\mathfrak{u}(1)\\
\text{ad}(\mathfrak{su}(2))  &  \rightarrow1_{0}+1_{2}+1_{-2}.
\end{align}
Recall that in this abelian example, the gauge field $A=0$, and the Higgs
field $\Phi=i\phi$ is given by:%
\begin{equation}
\Phi=\left[
\begin{array}
[c]{cc}%
df & \\
& -df
\end{array}
\right]
\end{equation}
with:
\begin{equation}
f=z\overline{z}-2t^{2}.
\end{equation}

Consider a candidate zero mode $\psi$ with charge $q=\pm2$ under the $\mathfrak{u}(1)$.
The analysis of zero modes for this case has already been studied in
references \cite{Acharya:1998pm, Acharya:2001gy, Pantev:2009de, Braun:2018vhk}. Writing out the
zero mode as a one-form on the three-manifold:
\begin{equation}
\psi=\psi_{z}dz+\psi_{\overline{z}}d\overline{z}+\psi_{t}dt,
\end{equation}
the zero mode equations are:
\begin{align}
0 &= \partial_{z}{\psi}_{\overline{z}}-\partial_{\overline{z}}{\psi}_{z}-qz{\psi
}_{z}+q{\psi}_{\overline{z}}  \\
0 &= \partial_{z}{\psi}_{t}-\partial_{t}{\psi}_{z}+qz{\psi}_{t}-4qt{\psi}_{z}  \\
0 &= \partial_{\overline{z}}{\psi}_{t}-\partial_{t}{\psi}_{\overline{z}}%
+q\overline{z}{\psi}_{t}-4qt{\psi}_{\overline{z}}  \\
0 &= g^{z\overline{z}}\left(  \partial_{z}{\psi}_{\overline{z}}-q\overline{z}{\psi
}_{\overline{z}}\right)  +g^{\overline{z}z}\left(  \partial_{\overline{z}%
}{\psi}_{z}-qz{\psi}_{z}\right)  +g^{tt}\left(  \partial_{t}{\psi}%
_{t}+4qt{\psi}_{t}\right)  .
\end{align}
We note that in the neighborhood where $t=0$, we must keep ${\psi}_{t}$
non-zero. To see why, observe that if it were zero, then there is a coupled
differential equation for ${\psi}_{z}$ and ${\psi}_{\overline{z}}$ which does
not produce a normalizable solution. Since perturbations involving ${\psi}%
_{t}$ do not really alter this conclusion, we see that a more sensible
starting point is to take ${\psi}_{t}$ non-zero and ${\psi}_{z}={\psi
}_{\overline{z}}=0$. In this case, the zero mode equations collapse to:%
\begin{align}
\partial_{z}{\psi}_{t}+qz{\psi}_{t}  &  =0\\
\partial_{\overline{z}}{\psi}_{t}+q\overline{z}{\psi}_{t}  &  =0\\
\partial_{t}{\psi}_{t}+4qt{\psi}_{t}  &  =0,
\end{align}
which is solved by:%
\begin{equation}
{\psi}_{t}\sim\exp(-q(z\overline{z}+t^{2})),
\end{equation}
so we get a normalizable zero mode for $q=+2$ but not for $q=-2$.

\subsection{Second Abelian Example \label{ssec:ABELIANII}}

As another example using a similar breaking pattern, we next consider the zero
modes in the background introduced in Section \ref{ssec:CODIM6}. In this
example we took the gauge group to be $SU(2)$ and its adjoint representation
decomposes as:%
\begin{align}
\mathfrak{su}(2)  &  \supset\mathfrak{u}(1)\\
\text{ad}(\mathfrak{su}(2))  &  \rightarrow1_{0}+1_{2}+1_{-2}.
\end{align}%
\begin{equation}
\mathcal{A}=\left[
\begin{array}
[c]{cc}%
\beta & \\
& -\beta
\end{array}
\right]  \,,
\end{equation}
with:%
\begin{equation}
\beta=zdz+\overline{z}d\overline{z}=2xdx-2ydy.
\end{equation}
The corresponding Higgs field in this case is:%
\begin{equation}
\Phi=\left[
\begin{array}
[c]{cc}%
df & \\
& -df
\end{array}
\right]
\end{equation}
with:
\begin{equation}
f=\frac{1}{2}z^{2}+\frac{1}{2}\overline{z}^{2}.
\end{equation}
By inspection, there is no $t$-dependence in the background fields, so we can
at best expect a localized mode in codimension six in the local $G_{2}$ background.

Consider a candidate zero mode $\psi$ with charge $q=\pm2$ under the $\mathfrak{u}(1)$.
The analysis of zero modes for this case is basically that already studied in
much of the F-theory literature for 6D theories, so we can borrow much of the
analysis from this work. Writing out the zero mode as a one-form on the
three-manifold:
\begin{equation}
\psi=\psi_{z}dz+\psi_{\overline{z}}d\overline{z}+\psi_{t}dt,
\end{equation}
the zero mode equations are:
\begin{align}
0 &= \partial_{z}{\psi}_{\overline{z}}-\partial_{\overline{z}}{\psi}_{z}%
-q\overline{z}{\psi}_{z}+qz{\psi}_{\overline{z}}  \\
0 &= \partial_{z}{\psi}_{t}-\partial_{t}{\psi}_{z}+qz{\psi}_{t}  \\
0 &= \partial_{\overline{z}}{\psi}_{t}-\partial_{t}{\psi}_{\overline{z}}+qz{\psi
}_{t}  \\
0 &= g^{z\overline{z}}\left(  \partial_{z}{\psi}_{\overline{z}}-qz{\psi}%
_{\overline{z}}\right)  +g^{\overline{z}z}\left(  \partial_{\overline{z}}%
{\psi}_{z}-q\overline{z}{\psi}_{z}\right)  +g^{tt}\left(  \partial_{t}{\psi
}_{t}\right)  .
\end{align}
In this case, we do not expect any $t$ dependence, so we can set
${\psi}_{t}=0$ as well. The system of equations now becomes:%
\begin{align}
\partial_{z}{\psi}_{\overline{z}}+qz{\psi}_{\overline{z}}  &  =\partial
_{\overline{z}}{\psi}_{z}+q\overline{z}{\psi}_{z}\\
\partial_{z}{\psi}_{\overline{z}}-qz{\psi}_{\overline{z}}  &  =-\partial
_{\overline{z}}{\psi}_{z}+q\overline{z}{\psi}_{z},
\end{align}
or equivalently:%
\begin{align}
\partial_{z}{\psi}_{\overline{z}}-q\overline{z}{\psi}_{z}  &  =0\\
\partial_{\overline{z}}{\psi}_{z}-qz{\psi}_{\overline{z}}  &  =0,
\end{align}
which has a normalizable solution which depends on the sign of $q$:%
\begin{align}
q  &  >0:{\psi}_{z}=-{\psi}_{\overline{z}}\sim\exp(-qz\overline{z}%
).\label{qpos}\\
q  &  <0:{\psi}_{z}=+{\psi}_{\overline{z}}\sim\exp(+qz\overline{z}).
\label{qneg}%
\end{align}
The presence of two solutions is of course consistent with the fact that in
this background, we have actually produced a 5D hypermultiplet.

Consider next the effect of perturbing our background via:
\begin{equation}
f_{\text{new}}=f + f_{\text{pert}}.
\end{equation}
The main point is that if $ f_{\text{pert}}$ has non-trivial
$t$-dependence, and is sufficiently small, then it cannot change the sign of
the Hessian in the original $(x,y)$ directions. Instead, the overall sign of
$\partial^{2} f_{\text{pert}} /\partial t^{2}$ will determine which of
the two components in lines (\ref{qpos}) and (\ref{qneg}) will actually
survive as a normalizable mode. That we can retain at most one normalizable mode
follows from the fact that we have a first order differential equation in the
$t$-variable.

\subsection{Fibering a Hitchin Solution}

The above example illustrates a far more general point which we can use to
generate a broad class of localized zero modes, even in the presence of
fluxes. Returning to our discussion in section \ref{ssec:HITPERT}, we start with a
three-manifold $M=\Sigma\times I$, and a solution to the Hitchin system on
$\Sigma$ given by $\mathcal{A}_{\text{H}}$ and then perturb it to a new solution:
\begin{equation}
\mathcal{A}_{\text{new}}=\mathcal{A}_{\text{H}}+ \mathcal{A}_{\text{pert}}.
\end{equation}
If we just confine our attention to zero modes on the curve, we observe that
we get a 5D hypermultiplet. In particular, we can take a T-brane background on
$\Sigma$ of the sort already considered in references \cite{Anderson:2013rka, Anderson:2017rpr}.
Perturbing this background further to have
non-trivial $t$-dependence, the discussion of section \ref{ssec:ABELIANII}
generalizes, and we obtain a 4D chiral multiplet. Observe that because we are
treating $\mathcal{A}_{\text{pert}}$ as a small perturbation, it does
not affect (to leading order) the localization on the curve $\Sigma$.
Additionally, the zero mode will have a very broad profile (though still
normalizable) on the interval $I$.

\subsection{T-Brane Example \label{ssec:TBRANEexample}}

We now turn to the background solution of Section \ref{sec:nabackground} and
analyze its zero mode content. As we already remarked there, an interesting
feature of this background is that the presentation of the spectral equation,
at least in holomorphic coordinates, appears to ``hide'' the appearance of the
singularity enhancement. This means in particular that even though the
geometry may appear to host a codimension six singularity, a localized chiral
mode may still be present.

We start by looking at the explicit form of the wave functions. To simplify
the analysis we temporarily set $f(z,\bar{z})=0$. Looking at the decomposition
$\mathfrak{su}(3)\rightarrow \mathfrak{su}(2)\times \mathfrak{u}(1)$ and taking the modes $\psi$ in the
$\mathbf{2}_{1}$ representation, we choose an ansatz of the form:
\begin{equation}
\psi=e^{-\frac{\kappa}{2}t^{2}}\left(  \left[
\begin{array}
[c]{c}%
\tau_{1}(z,\bar{z})\\
i\alpha\beta(z,\bar{z})
\end{array}
\right]  dz+\left[
\begin{array}
[c]{c}%
\alpha\beta(z,\bar{z})\\
\tau_{2}(z,\bar{z})
\end{array}
\right]  d\bar{z}\right)  \,.
\end{equation}
Here, $\alpha$ is a constant parameter which determines the norm of the wave function. We have three non-trivial functions
$\tau_1, \tau_2, \beta$ which we must determine. These functions satisfy the differential equations:
\begin{align}
\frac{\alpha}{4}\kappa\left(  \bar{z}+z\right)  \beta\left(  z,\bar{z}\right)
+\partial_{\bar{z}}\tau_{1}\left(  z,\bar{z}\right)  -i\varepsilon\tau
_{2}\left(  z,\bar{z}\right)   &  =0\,,\\
-\frac{i\alpha}{4}\kappa\left(  \bar{z}+z\right)  \beta\left(  z,\bar
{z}\right)  +\partial_{z}\tau_{2}{}\left(  z,\bar{z}\right)  +i\varepsilon
\tau_{1}\left(  z,\bar{z}\right)   &  =0\,,\\
\alpha\partial_{\bar{z}}\beta\left(  z,\bar{z}\right)  +\alpha\upsilon
z\beta\left(  z,\bar{z}\right)  +\frac{i\kappa}{4}\left(  \bar{z}+z\right)
\tau_{2}\left(  z,\bar{z}\right)   &  =0\,,\\
\alpha\partial_{z}\beta\left(  z,\bar{z}\right)  +\alpha\upsilon\bar{z}%
\beta\left(  z,\bar{z}\right)  +\frac{\kappa}{4}\left(  \bar{z}+z\right)
\tau_{1}\left(  z,\bar{z}\right)   &  =0\,.
\end{align}
While we are not able to find a solution to the full system we can take
$\kappa\ll1$. The zeroth order in $\kappa$ equations imply that $\tau_{1}%
=\tau_{2}=0$ and $\beta=e^{-\upsilon z\bar{z}}$. At first order in $\kappa$
the equations become
\begin{align}
\frac{\alpha}{4}\kappa\left(  \bar{z}+z\right)  e^{-\upsilon z\bar{z}%
}+\partial_{\bar{z}}\tau_{1}\left(  z,\bar{z}\right)  -i\varepsilon\tau
_{2}\left(  z,\bar{z}\right)   &  =0\,,\\
-\frac{i\alpha}{4}\kappa\left(  \bar{z}+z\right)  e^{-\upsilon z\bar{z}%
}+\partial_{z}\tau_{2}{}\left(  z,\bar{z}\right)  +i\varepsilon\tau_{1}\left(
z,\bar{z}\right)   &  =0\,.
\end{align}
For $\varepsilon=0$ it is possible to solve these equations and we get
\begin{align}
\beta(z,\bar{z})  &  =e^{-z\bar{z}}+\mathcal{O}(\kappa^{2})\,,\\
\tau_{1}(z,\bar{z})  &  =i\alpha\kappa\left[  e^{-z\bar{z}}\frac{1+z^{2}%
+z\bar{z}}{4z^{2}}-\frac{1}{4z^{2}}\right]  +\mathcal{O}(\kappa^{2})\,,\\
\tau_{2}(z,\bar{z})  &  =\alpha\kappa\left[  e^{-z\bar{z}}\frac{1+\bar{z}%
^{2}+z\bar{z}}{4\bar{z}^{2}}-\frac{1}{4\bar{z}^{2}}\right]  +\mathcal{O}%
(\kappa^{2})\,.
\end{align}
We see that the wave functions are fully localized. The function $h$ was not
capable of producing a chiral mode since it has a zero eigenvalue in its Hessian. Instead,
this contribution would only lead to a codimension six
singularity in the local geometry. The addition of the Hitchin system gives a mode that is localized
at a point. One interesting feature of this background is that the Hitchin
system provides a deformation that is invisible in the geometry due to the
fact that it is nilpotent when $\varepsilon = 0$. This implies that, whereas the geometry might seem
to host a codimension six singularity and therefore non-chiral matter,
the presence of a T-brane in the system can allow the system to still support a localized chiral mode.

To further confirm the existence of a zero mode we follow the
strategy outlined in section \ref{sec:cohom}. The mode $\psi_{0}$ that
provides a solution to the F-term equations is in this case
\begin{equation}
\psi_{0}=e^{\frac{1}{2}f(z,\bar{z})}\,e^{-\frac{\kappa t^{2}}{2}+\frac{\kappa
}{8}(z+\bar{z})^{2}}\,\eta(z,\bar{z})\left(  \left[
\begin{array}
[c]{c}%
0\\
-i\alpha
\end{array}
\right]  dz+\left[
\begin{array}
[c]{c}%
\alpha\\
0
\end{array}
\right]  d\bar{z}\right)  \,,
\end{equation}
where $\eta(z,\bar{z})$ depends only on the combination $r^{2}\equiv z\bar{z}$
and satisfies the differential equation
\begin{equation}
\eta^{\prime}(r) = -2vr \eta(r) e^{-f(r)}\,.
\end{equation}
For example by taking only the zeroth order in the expansion of $f(r)$
near $r=0$ we find that
\begin{equation}
\eta(r)=e^{-\upsilon r^{2}e^{-f(0)}}\,,
\end{equation}
which shows that $\psi_{0}$ is indeed normalizable provided that $\upsilon$ is
sufficiently large compared to $\kappa$.

Summarizing, we see that although the T-brane background with $\varepsilon = 0$ has a degenerate
spectral equation in the holomorphic geometry (as per our discussion in section
\ref{sec:nabackground}), we also see that there is a localized chiral zero
mode. Although we have only given an approximate characterization of the exact
form of this wave function, we have demonstrated existence by showing that
there are nearby ``trial wave functions'' which provide good approximations.
Finally, it is straightforward to generalize this class of examples to cover
an analysis of zero modes for more general T-brane backgrounds of the sort
discussed in section \ref{ssec:more}. All that is really required is a
suitable embedding $\mathfrak{su}(2) \times\mathfrak{u}(1) \rightarrow
\mathfrak{g}_{\mathbb{C}}$.

\section{Algebraic Approach} \label{sec:ALGEBRAIC}

As explained in previous sections, the procedure for determining the appearance
of localized chiral matter splits into two steps.
First of all, we must ensure that we can produce
an actual background which solves the PW equations, a system of
second order partial differential equations.
Secondly, the fluctuations around a given background are also governed by
second order partial differential equation.

Our aim in this section will be to assume the existence of a
consistent background and then develop some algebraic methods to analyze
the local behavior of zero modes. Locally, at least, solutions to the F-term equations
motion can always be written as:
\begin{equation}
\mathcal{A} = g^{-1} dg
\end{equation}
for suitable $g(x)$ a $G_{\mathbb{C}}$ valued function. We can then
either attempt to solve the D-term constraints (in unitary gauge), or equivalently, determine
constraints on $g(x)$ such that we have a stable background solution (in complexified gauge).
Our task will be to determine the zero mode spectrum for a given choice of $g(x)$.

We present our prescription for extracting the zero mode spectrum from a given
background as a conjectural proposal motivated
by physical considerations, that is, we shall
not give a complete derivation of this prescription. Instead, we shall present
a sketch of how to derive these results and show that it correctly computes
the localized zero mode spectrum in all the examples encountered previously.

We first study the spectral equations for the
Higgs field and show that compared with the case of intersecting seven-branes in
F-theory (as studied for example in \cite{Hayashi:2009ge, Donagi:2009ra, Marsano:2009gv, Blumenhagen:2009yv}),
the real structure of the Higgs field complicates the use of such methods. There is, however,
a remnant of holomorphic structure which can be fruitfully used to extract the relevant structure of localized zero modes,
and this leads us to our general prescription for extracting 4D chiral matter. First, we take a local solution to the background
equations of motion and show that the destabilizing wall in the moduli space is captured by a local Hitchin system on a 2D subspace.
There is a natural holomorphic structure for our Higgs field here as well as a corresponding 5D hypermultiplet.
Perturbations away from this solution produce localized 4D chiral matter. With
this machinery in place, we then revisit the examples previously discussed,
showing that we indeed correctly capture the local profile of zero modes in
such backgrounds.

\subsection{Spectral Methods and Their Limitations}

In this section we discuss some limitations of the spectral equation(s) of the Higgs field
as a tool in understanding the structure of localized zero modes. The PW system tells us about
intersecting six-branes in the local geometry $T^{\ast}M$. For ease of
exposition, we shall often specialize to the case where the only non-zero
values of the Higgs field are in an $\mathfrak{su}(N)$ subalgebra, and present
the spectral equation in the fundamental representation:
\begin{equation}
\det\left(  u_{i}\mathbb{I}_{N}-\Phi_{i}\right)  =0,
\end{equation}
where the $u_{i}$ denote directions in the cotangent bundle $T^{\ast}M$
with zero section $\{u_{i} = 0 \}$ given by $M$.
This specifies the constraint equations in $T^{\ast}M$, and thus yields a
three-dimensional spectral manifold in the ambient geometry.

We are interested in situations in which a non-trivial flux may be present,
so the components of the Higgs field will not commute, i.e. $[\Phi_{i},\Phi_{j}]\neq0$.
Even though we cannot simultaneously diagonalize
the components of the Higgs field, observe that since the spectral
equations are gauge invariant, we are still free to write:
\begin{equation}
\det(u_{j}\mathbb{I}_{N}-\Phi_{j})=\prod_{I_j=1}^{N}\left(  u_{j}-\lambda
_{j}^{I_j}\right)  =0,
\end{equation}
for $j=1,2,3$ the three components of the one-form. Here,
we also introduced an index $I_j = 1,...,N$ which runs over the eigenvalues of each
component of the Higgs field. Since we do not want to assume the components of the
Higgs fields simultaneously commute, we must separately index the eigenvalues for
each value of $j$. In other words, there is no well-defined spectral line bundle
when the components of the Higgs field do not commute.

The complications arise when we attempt to extract the zero mode content from such a background.
Mimicking what happens in the abelian case, we first write down the intersections of different
sheets in the spectral equations:
\begin{align}
 u_{1} - \lambda_{1}^{I_1} & = u_{1} - \lambda_{1}^{J_1}\\
 u_{2} - \lambda_{2}^{I_2} & = u_{2} - \lambda_{2}^{J_2} \\
 u_{3} - \lambda_{3}^{I_3} & = u_{3} - \lambda_{3}^{J_3}.
\end{align}
Here then, is the issue: since we cannot simultaneously diagonalize the three components of the Higgs field, we have
no way of extracting the locus where a broken symmetry is restored at points of the three-manifold. Said differently,
in the case of a T-brane configuration there is no canonical way to order the $I_j$ and $J_j$ for
different values of $j$.

To address this shortcoming we will need to introduce additional structure into the local system. We will shortly see that
there is a well-motivated prescription which allows us to build a (nearly) holomorphic Higgs field, and thus apply spectral
equation methods.

\subsection{Local Matter Ring}

We now turn to an algebraic characterization of zero modes in a local patch. We shall
emphasize the existence of these zero modes rather than the explicit profile
of the wave function, as the latter requires solving an explicit second order
partial differential equation. This means that we primarily emphasize
purely holomorphic F-term data and suitable stability conditions.

Our analysis is similar in spirit to that carried out in reference
\cite{Cecotti:2010bp} (see also \cite{Donagi:2003hh, Cecotti:2009zf}) though
there are some differences. There, the essential picture involved determining
the spectrum of B-branes in a local Calabi--Yau. Here, we are studying a related
question but for A-branes in the local Calabi--Yau $T^{\ast}M$.

Zero modes are captured by linearized perturbations in the F-term equations of
motion modulo complexified gauge transformations:%
\begin{equation}
D_{\mathcal{A}}\psi=0\text{ \ \ modulo \ \ }\psi\simeq\psi+D_{\mathcal{A}}%
\chi,
\end{equation}
with $\chi$ an adjoint valued zero-form.

Now, since we are working in a local patch, all of the possible backgrounds
are determined by a choice of $G_{\mathbb{C}}$ valued function $g(x)$ via:
\begin{equation}
\mathcal{A}=g^{-1}dg.
\end{equation}
In this local patch, gauge transformations can be viewed as right multiplication by
$G_{\mathbb{C}}$ valued functions $\alpha(x)$. To see why,
observe that under right multiplication with $g_{\text{new}%
}\mathcal{=}g\alpha$, we have:%
\begin{equation}\label{newcalA}
\mathcal{A}_{\text{new}}=g_{\text{new}}^{-1}dg_{\text{new}}=(g\alpha
)^{-1}d(g\alpha)=\alpha^{-1}\mathcal{A\alpha+\alpha}^{-1}d\alpha,
\end{equation}
namely a complexified gauge transformation. An important subtlety with this
procedure is we must suitably limit the space of functions $\alpha$ to be only
those which respect a suitable notion of stability, which necessarily excludes
some candidate gauge transformations. For example, consider
the Morse function $f=x^{2}+y^{2}-2t^{2}$ and use a formal gauge
transformation to flip the signature of the Hessian. Clearly, such gauge
transformations must be excluded.

Let us next turn to candidate zero modes. These are captured by small
perturbations to a given background determined by a choice of $g(x)$. In other
words, perturbations are specified by functions taking values in the Lie algebra, i.e.:%
\begin{equation}
\text{Candidate Zero Modes: }\mathcal{O}\otimes\mathfrak{g}_{\mathbb{C}},
\end{equation}
where $\mathcal{O}$ is the local ring of complex functions. This amounts to
the special case:%

\begin{equation}
g_{\text{new}}=g(1+h+...),
\end{equation}
where $h$ is an infinitesimal Lie algebra valued function $h \in \mathcal{O} \otimes \mathfrak{g}_{\mathbb{C}}$.
The perturbed connection is of the form:%
\begin{equation}
\mathcal{A}_{\text{new}}=\mathcal{A}+D_{\mathcal{A}}h,
\end{equation}
and a zero mode is given by the formal expression:%
\begin{equation}
\psi=D_{\mathcal{A}}h.
\end{equation}
As we remarked previously, even though this may appear to be a
\textquotedblleft pure gauge\textquotedblright\ configuration, a suitable
notion of stability will enforce the condition that such solutions cannot be
gauged away.

Indeed, even though we can formally consider general complexified gauge transformations, these transformations can
impact the asymptotic profile of the fields at the boundary of a patch. Since this data on the patch must be held fixed,
this necessarily excludes some complexified gauge transformations. In practical terms, we see the effects of passing
from a stable solution to an unstable one precisely through a possible jump in the chirality of the spectrum in a given patch.

How then, shall we proceed in extracting the zero mode spectrum from a
candidate background? We expect that zero modes are captured by an annihilator
condition involving the Higgs field $\Phi=i\phi$. Our proposal is to focus on
taking complexified gauge transformations to be \textquotedblleft as close to
unstable\textquotedblright\ as possible. By this we mean that if we make the
gauge transformation any larger, we would flip the chirality of the candidate
localized zero mode, which would clearly violate any reasonable notion of
stability in a local patch. This also means that there is a special class of
singular complexified gauge transformations which will produce a non-chiral
spectrum, i.e. a chiral and anti-chiral pair of zero modes. It is this special
case on which we choose to focus.

We start with one of these
non-chiral solutions, and then perturb it back to a chiral solution:%
\begin{equation}
\mathcal{A}= \mathcal{A}_{\text{non-chiral}}+\mathcal{A}_{\text{pert}}\text{.}%
\end{equation}
As explained near equation (\ref{newcalA}), all local complexified gauge transformations can be viewed as right multiplication by a
$G_{\mathbb{C}}$-valued gauge parameter. Consequently, we can start with the case of $\mathcal{A}_{\text{non-chiral}} = g_{\mathrm{non-chiral}}^{-1} g_{\mathrm{non-chiral}}$ and multiply to $g_{\mathrm{non-chiral}} \rightarrow g_{\mathrm{non-chiral}} \alpha $.
The transformed value of the gauge parameter will in general alter the profile of the fields at the boundary of the patch. Changing such asymptotic profiles means in turn that we have jumped from one stable solution to another. As we have remarked previously, such a jump
generically also leads to a change in the chirality of the spectrum, a feature we will explicitly verify in examples.

Now, by definition, $\mathcal{A}_{\text{non-chiral}}$ only has non-trivial support
on two out of the three directions of the local coordinates, which we refer to
as $z$ and $\overline{z}$. We refer to the third direction as $t$. We
interpret $z$ and $\overline{z}$ as coordinates for a local patch of a Riemann
surface $\Sigma$, and $\mathcal{A}_{\text{non-chiral}}$ as a complexified flat
connection (which will shortly be perturbed). This means that we can split up
our analysis of localization into a contribution on $\Sigma$ and then a
perturbation normal to $\Sigma$ inside of $M$. We note that in unitary gauge,
the analogous statement is that we are free to take the $g^{tt}\rightarrow0$
limit in the D-term constraint, and then perturb away from the singular limit.

\begin{figure}[t]
\begin{center}
\includegraphics[trim={0cm 0cm 0cm 0cm},clip,scale=0.5]{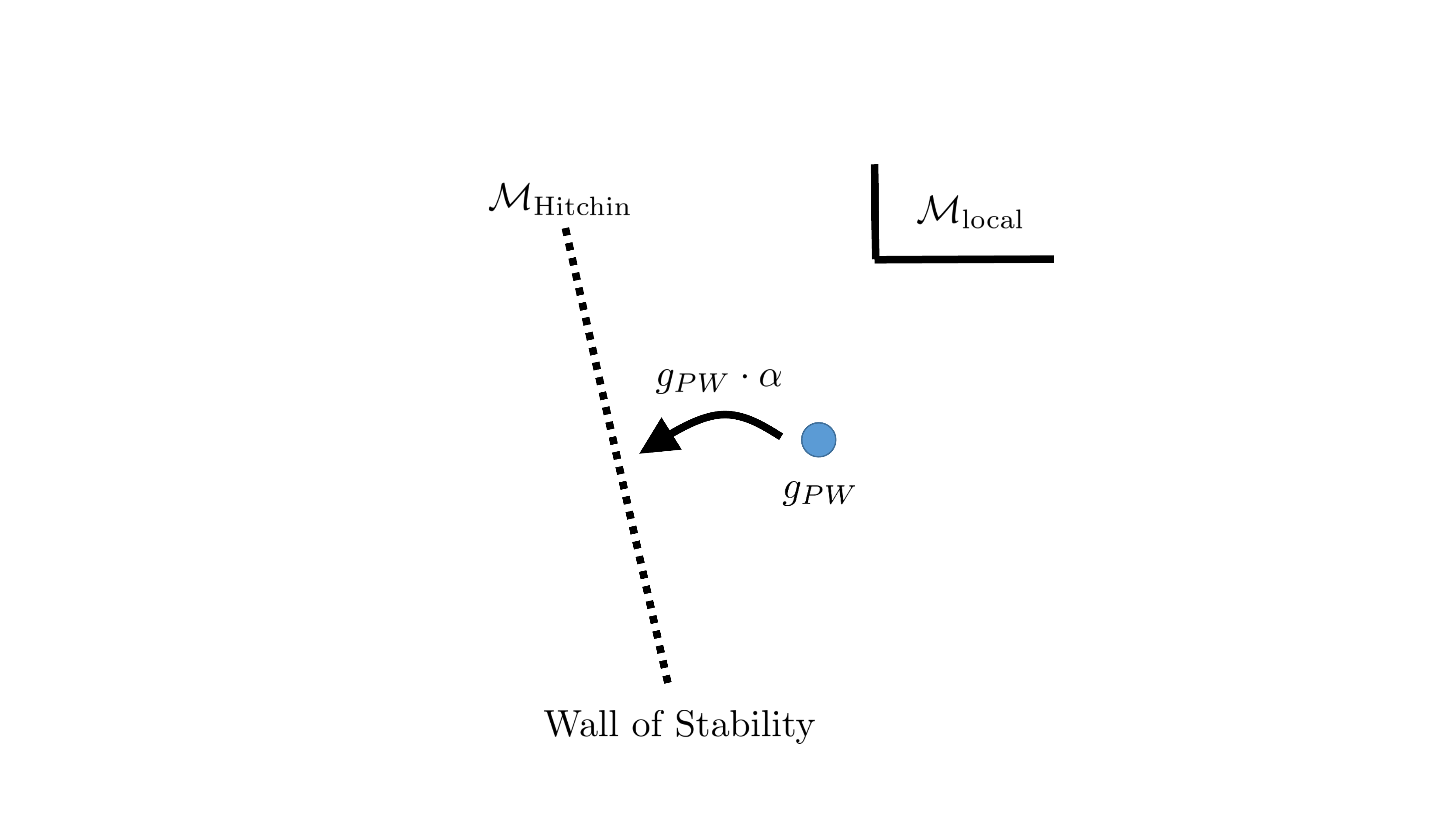}
\end{center}
\caption{Depiction of the moduli space of solutions to the PW
system in a local patch. Each solution is captured by a $G_{\mathbb{C}}$ valued
function $g_{PW}(x)$, which determines a complexified flat
connection $\mathcal{A}_{PW}=g_{PW}^{-1}dg_{PW}$.
Complexified gauge transformations correspond to right multiplication by $G_{\mathbb{C}}$
valued functions $\alpha(x)$ subject to
suitable stability conditions. The walls of stability are captured by a local
Hitchin system defined on a 2D subspace inside the three-manifold.
Perturbations away from this solution define the gauge equivalence classes of
solutions in the PW system.}%
\label{fig:STAB}%
\end{figure}

Now, on $\Sigma$, the background $\mathcal{A}_{\text{non-chiral}}$ determines
a solution to the Hitchin system (it is a complexified flat connection), and
we can first ask whether we can find a localized zero mode here. In fact, this
is precisely what was already worked out in reference \cite{Cecotti:2010bp} in
the related case of intersecting seven-branes wrapped on K\"{a}hler surfaces.
The present case is a mild adaptation of that, so we just summarize the main
idea. First, we introduce an adjoint-valued $(1,0)$ form Higgs field $\Phi
_{z}$. Then, localized matter on the Hitchin curve is obtained from the
quotient:%
\begin{equation}
\text{5D Zero Mode on }\Sigma\simeq\frac{\mathcal{O}_{\Sigma}^{\text{hol}%
}\otimes\mathfrak{g}_{\mathbb{C}}}{\left\langle \ker\text{ad}_{\Phi_{z}%
}\right\rangle },
\end{equation}
where here, $\mathcal{O}_{\Sigma}^{\text{hol}}\simeq\mathbb{C}[[z]]$ is a
formal power series in the local holomorphic coordinate. In the full system on
a three-manfold, we will need to work in terms of the power series
$\mathbb{C}[[z,\overline{z}]]\simeq\mathbb{C}[[x,y]]$ since $\Phi_{z}$ will
typically not depend on just $z=x+iy$. When we do this, we are taking an
annihilator condition such as:%
\begin{equation}
h+z\chi=0,
\end{equation}
for $h$ our zero mode and $\chi$ our gauge redundancy parameter. In some
non-abelian configurations, it is convenient to instead work with real
coordinates $x$ and $y$. Owing to the holomorphic structure of our system, we
can equivalently treat such conditions as split up into two separate
conditions since we can always restrict $h$ to be real, and separately impose
an annihilator constraint from $x$ and $y$.

Perturbing this 5D\ zero mode, we then reach a 4D chiral mode. From the
argument just presented, we can take a family of embeddings of our Riemann
surface in the $t$ direction, and by a choice of unitary gauge transformation,
we can, without loss of generality, set $A_{t}=0$ so that the complexified
connection in this direction is just the Higgs field component $\Phi_{t}$.
This also means that the study of localized zero modes reduces to determining
the adjoint action by just $\Phi_{t}$. Since the ring of perturbations is now
in $\mathcal{O}\otimes\mathfrak{g}_{\mathbb{C}}\simeq\mathbb{C}%
[[x,y,t]]\otimes\mathfrak{g}_{\mathbb{C}}\simeq\mathbb{C}[[z,\overline
{z},t]]\otimes\mathfrak{g}_{\mathbb{C}}$, we reach our proposal for
4D\ localized zero modes:%
\begin{equation}
\text{4D Local Zero Modes}\simeq\frac{\mathcal{O}\otimes\mathfrak{g}%
_{\mathbb{C}}}{\left\langle \ker\text{ad}_{\Phi_{z}},\ker\text{ad}_{\Phi_{t}%
}\right\rangle },
\end{equation}
that is, we quotient by the adjoint action of both $\Phi_{t}$ and $\Phi_{z}$.
Observe, however, that we are not quotienting by $\Phi_{\overline{z}}$.
Indeed, this is not what we do to extract the 5D\ zero modes anyway.

Our proposal can now be summarized as follows. We observe that solutions to
the Hitchin system on a local 2D subspace serve as walls of stability,
and perturbations away from these solutions produce
the generic structure of solutions to our backgrounds. Indeed, we have
presented a general argument that if we attempt to perform a formal gauge
transformation which moves through such a wall, the spectrum of chiral zero
modes will reverse signs. This can happen in a local patch, but clearly
delineates the ``boundary behavior'' in the moduli space of the PW
system (see figure \ref{fig:STAB}).

In practice, this also means we will need to find a suitable coordinate system
such that a Hitchin system can be defined on the subspace in the first place.
We expect that this is generically possible because we can also view this
procedure of taking complexified gauge transformations as alternatively
varying the background metric on the three-manifold until we produce a
(singular) limit where the metric collapses to that on a patch of a Riemann
surface. In what follows we shall operate under the assumption that all
walls of stability in the PW system are captured by a Hitchin system and then
extract the localized zero mode spectrum by perturbing away from these walls.

Now, a pleasant feature of the Hitchin system is that there is a spectral equation in terms of a holomorphic
section of a bundle $\varphi_{\text{hol}}$. For example, in the case $G = SU(N)$, the spectral equation
in the fundamental representation has the form:
\begin{equation}
\det ( u_z \mathbb{I}_{N} - \varphi_{\text{hol}}) = 0
\end{equation}
which is a hypersurface in $T^{\ast} \Sigma$ with $\Sigma$ our ``local Hitchin system curve.'' Since this is a
single holomorphic equation, we can use spectral equation methods to often extract the zero mode content in such cases.
Of course, it is also well known that this method can also fail in some T-brane configurations
when $\varphi_{\text{hol}}$ is nilpotent \cite{Cecotti:2010bp}.

Our plan in the remainder of this section will be to revisit the examples of zero modes
treated in section \ref{sec:MATTER}. It will hopefully become apparent that
the algebraic approach presented here provides a computationally powerful way to characterize
zero mode localization in general fluxed backgrounds.

\subsection{Examples Revisited}

In section \ref{sec:MATTER} we showed how to read off the explicit wave
function profile for zero mode solutions in a given background. In this
section we use algebraic methods to at least argue for the existence and
location of zero modes. We proceed by considering each of the previously
studied examples. For ease of exposition, we shall also often set unimportant
factors to unity.

\subsubsection{Abelian Example}

We now analyze the zero mode content of the background obtained in section
\ref{sec:abackground} with Higgs field:
\begin{equation}
\Phi_{z}\sim\left[
\begin{array}
[c]{cc}%
\overline{z} & \\
& -\overline{z}%
\end{array}
\right]  \,,\text{ \ \ }\Phi_{t}\sim\left[
\begin{array}
[c]{cc}%
-4t & \\
& 4t
\end{array}
\right]  .
\end{equation}

To analyze the zero mode content, we first take a complexified
gauge transformation to make the background \textquotedblleft as close as
possible\textquotedblright\ to that of a Hitchin system on a (non-compact)
curve. The first subtlety we face is that the Hessian for $f=z\overline
{z}-2t^{2}=x^{2}+y^{2}-2t^{2}$ has signature $(+,+,-)$ in the $(x,y,t)$
coordinates.

Since we are interested in a stable solution to the Hitchin system equations, we need to pick local coordinates which retain a single $+$ and a single $-$ in the Hessian. Consequently, the coordinates for the local Hitchin system must
involve the $t$ coordinate (since this is the only $-$ eigenvalue in the Hessian), so we introduce shifted variables: $z^{\prime
}=x+it$, $\overline{z}^{\prime}=x-it$, $t^{\prime}=y$, with the 2D subspace
for our Hitchin system on $z^{\prime}$ and $\overline{z}^{\prime}$. Observe,
however, that since we are discussing an abelian example, these distinctions
do not really matter and we can just analyze the annihilator conditions for
$\Phi_{x}$, $\Phi_{y}$ and $\Phi_{t}$. We caution that this step is in general
not valid and only holds for abelian examples with no flux.

We next determine the structure of the local ring of zero modes in this case.
Without loss of generality, we take a zero mode $h$ of charge $q=2$. Along the
$z^{\prime}$ direction, the gauge condition is:%
\begin{equation}
h+q(1-4)z^{\prime}\chi=0,
\end{equation}
so we can gauge away any non-trivial $z^{\prime}$ dependence. This means our
5D\ zero mode is localized along $\mathbb{C}[[z^{\prime}]]/\left\langle
z^{\prime}\right\rangle \simeq\mathbb{C}$. Next, we turn to the $t^{\prime}$
annihilator condition:%
\begin{equation}
h+qt^{\prime}\eta=0,
\end{equation}
for $\eta$ a gauge parameter.\footnote{One might ask why we are introducing
$\eta$ another gauge parameter. The point is that we have a separate adjoint action coming
from $\Phi_{t}$ to consider.}
So, we conclude that the zero modes are captured by the ring:%
\begin{equation}
\text{4D Local Zero Mode Space: \ \ }\frac{\mathbb{C}[[z^{\prime},\overline
{z}^{\prime},t^{\prime}]]}{\left\langle z^{\prime},\overline{z}^{\prime
},t^{\prime}\right\rangle }\simeq\mathbb{C},
\end{equation}
i.e. a single 4D chiral multiplet localized at the origin $x=y=t=0$. Note that
we can also use our perspective on Hitchin systems as destabilizing walls to
tell us whether we produced a chiral or anti-chiral zero mode, i.e. based on
the sign of a perturbation away from the local system captured by the
coordinates $(z^{\prime},\overline{z}^{\prime})$.

\subsubsection{Second Abelian Example \label{ssec:ABELIANIIagain}}

As another example using a similar breaking pattern, we next consider the zero
modes in the background introduced in section \ref{ssec:CODIM6}. In this
example we took:%
\begin{equation}
\Phi_{z}\sim\left[
\begin{array}
[c]{cc}%
z & \\
& -z
\end{array}
\right]  \,,\text{ \ \ }\Phi_{t}\sim\left[
\begin{array}
[c]{cc}%
0 & \\
& 0
\end{array}
\right]  .
\end{equation}
Since there is no $t$-dependence in the background fields, we can at best
expect a localized mode in codimension six in the local $G_{2}$ background.

The spectral equation for $\Phi_{z}$ in the fundamental representation is:
\begin{equation}
u_{z}^{2}-z^{2}=0.
\end{equation}
Based on the form of this equation, we expect there to be a localized 5D
hypermultiplet at the common zero for the eigenvalues: $z=0$.

We next determine the structure of the local ring of zero modes in this case.
Without loss of generality, we take $q= \pm 2$ for our zero mode obtained from
adjoint breaking. A priori, the three components of the one-form are each
elements in the power series $\mathbb{C}[\left[  x,y,t\right]  ]$ where we
also use $z=x+iy$. Given $h$ a charge $q$ zero-form in $\mathcal{O}%
\simeq\mathbb{C}[\left[  x,y,t\right]  ]$, and $\chi$ a gauge parameter, the
algebraic annihilator condition is:%
\begin{equation}
h+qz\chi=0,
\end{equation}
so we conclude that the zero modes are captured by the ring:%
\begin{equation}
\text{5D Local Zero Mode Space: \ }\frac{\mathbb{C}[z]}{\left\langle
z\right\rangle }\simeq\mathbb{C},
\end{equation}
i.e. a single 5D hypermultiplet localized at the origin $x=y=0$.

Next consider perturbations to our background $\Phi$ in the $t$ direction:%
\begin{equation}
\Phi_{\text{new}}=\Phi+\Phi_{\text{pert}}.
\end{equation}
We assume that $\Phi_{\text{pert}}$ takes values in the same Cartan subalgebra
as $\Phi$. It is immediate that if the perturbations in the $z$ direction are
sufficiently small, the only change is in the $t$-dependence of a candidate
zero mode profile. This in turn means that for the most generic case where
$\Phi_{\text{pert}}$ contains a term proportional to $tdt$, that the 4D local
zero mode space is:%
\begin{equation}
\text{4D Local Zero Mode Space: \ \ }\frac{\mathbb{C}[[x,y,t]]}{\left\langle
x,y,t\right\rangle }\simeq\mathbb{C},
\end{equation}
as expected.

\subsubsection{Fibering a Hitchin Solution}

We now generalize the considerations of the above example to consider 5D
hypermultiplets which are localized on the $\Sigma$ factor of $M=\Sigma\times
I$. We start with a solution to the Hitchin system on $\Sigma$ given by
$\Phi_{\text{H}}$ and perturb it to generate a solution to the fluxed
PW\ equations:%
\begin{equation}
\Phi_{\text{new}}=\Phi_{\text{H}}+\Phi_{\text{pert}}.
\end{equation}
Provided $\Phi_{\text{pert}}$ is small, we can first analyze the zero mode
content on $\Sigma$, and then extend it to $M$. The annihilator conditions on
$\Sigma$ are basically the same as those used in reference
\cite{Cecotti:2010bp}, so we conclude that the algebraic analysis also
correctly captures the zero mode content in this more general situation
(almost by design).

\subsubsection{T-Brane Example}

As our final example, we turn to the fully non-abelian configuration of
section \ref{sec:nabackground} and use our algebraic methods to analyze the
zero mode content in this case as well. We first need to introduce a
convenient splitting of the PW\ Higgs field into one associated with a Hitchin
system, and then a perturbation away from this choice:%
\begin{equation}
\Phi_{\text{PW}}=\Phi_{\text{H}}+\Phi_{\text{pert}}\text{,}%
\end{equation}
where%
\begin{equation}
\Phi_{\text{H}}\text{,}\sim\left[
\begin{array}
[c]{ccc}
& \varepsilon dz & \\
zdz &  & \\
&  & 0
\end{array}
\right]  \,,\text{ \ \ }\Phi_{\text{pert}}\sim\left[
\begin{array}
[c]{ccc}%
xdx+tdt &  & \\
& xdx+tdt & \\
&  & -2(xdx+tdt)
\end{array}
\right]  .
\end{equation}
As a one-form, the perturbation points in the $d(x+t)$ direction.

We focus our attention on zero modes in the doublet representation under the
breaking pattern $\mathfrak{su}(3)\supset\mathfrak{su}(2)\times\mathfrak{u}%
(1)$, captured by the entries:%
\begin{equation}
\psi=\left[
\begin{array}
[c]{ccc}%
\mathcal{\ast} & \mathcal{\ast} & \psi^{+}\\
\mathcal{\ast} & \mathcal{\ast} & \psi^{-}\\
\ast & \ast & \ast
\end{array}
\right]  \,.
\end{equation}
Denoting by $h$ the corresponding doublet of localized fluctuations and $\chi$
the doublet of gauge parameters entering the annihilator equations, the zero
modes fill out entries in $\mathcal{O}\otimes\mathbb{C}^{2}$. In what follows,
we treat separately the cases $\varepsilon\neq0$ and $\varepsilon=0$.

Assuming  $\varepsilon\neq0$,  the annihilator conditions coming from
$\Phi_{\text{H}}$ are:%
\begin{equation}
\Phi_{\text{H}}:\left[
\begin{array}
[c]{c}%
h^{+}+\varepsilon\chi^{-}\\
h^{-}+z\chi^{+}%
\end{array}
\right]  =0.\label{hitchann}%
\end{equation}
We are then free to use our gauge parameter $\chi^{-}$ to eliminate $h^{+}$.
The second equation tells us:%
\begin{equation}
h^{-}+z\chi^{+}=0.
\end{equation}
so we get only a single zero mode trapped at the origin:%
\begin{equation}
\text{5D Local Zero Mode Space: \ \ }\frac{\mathbb{C}[[z]]}{\left\langle
z\right\rangle }\simeq\mathbb{C}.
\end{equation}

Consider next the annihilator conditions coming from $\Phi_{\text{pert}}$. As
a one-form, this points in the $d(x+t)$ direction, but since we have already
set fluctuations in the $x=0$ direction to zero, it is enough to analyze the
annihilator condition in just the $dt$ direction. Additionally, since we have
already gauge away $h^{+}$, we are left to analyze just $h^{-}$. We have:%
\begin{equation}
\Phi_{\text{pert}}:h^{-}+t\eta^{-}=0,
\end{equation}
from which it follows that we have a single localized zero mode at the origin:%
\begin{equation}
\text{4D Local Zero Mode Space }(\varepsilon\neq0)\text{: \ \ }\frac
{\mathbb{C}[[x,y,t]]}{\left\langle x,y,t\right\rangle }\simeq\mathbb{C}.
\end{equation}

We now turn to the case of $\varepsilon=0$. The analysis of the $h^{-}$ zero
mode is basically the same as the case with $\varepsilon\neq0$, so we have at
least a 4D\ local zero mode at $x=y=t=0$. Note, however, that now we cannot
gauge away $h^{+}$ since our gauge parameter $\chi^{-}$ no longer appears in
equation (\ref{hitchann}). So in this case, the algebraic approach can at best
point towards partial localization of a zero mode. The full analysis of the
system given earlier in section \ref{sec:MATTER} reveals that there is still
just a single 4D chiral multiplet localized at the origin, regardless of
whether $\varepsilon$ is zero or non-zero.

\section{Conclusions \label{sec:CONC}}

Compactifications of string / M- / F-theory on manifolds of special holonomy produce a wide variety of
novel physical systems. In this paper we have used methods from three- and
two-dimensional gauge theory to study a broad class of examples in which some
important features of the background are captured by non-abelian data, namely
T-brane configurations. We have shown that T-brane configurations in $G_{2}$
backgrounds are rather ubiquitous. Additionally, we have seen that the local
gauge theory description of $G_{2}$ backgrounds can be understood as a
deformation of Calabi--Yau threefolds fibered over an interval. In gauge theory
terms this is captured by a gradient flow equation in a deformation of a
Hitchin-like system on a Riemann surface. We have also shown how to
algebraically extract the zero mode content from these solutions. In the
remainder of this section we discuss some avenues of further investigation.

We have presented some evidence that the localized zero mode content can be
obtained from local algebraic conditions. It would of course be
desirable to have a more systematic derivation of this claim. A related
comment is that our analysis provides a potential way to characterize the
spectrum of bound states of A-branes in a local Calabi--Yau threefold.\ It
would be very interesting to develop this treatment further.

The general structure of fluxed non-abelian solutions involves fibering a
2D\ gauge theory over an interval to produce a 3D\ gauge theory with moduli
space matching onto that of a $G_{2}$ background. It is quite natural to
extend this procedure to consider 3D PW\ gauge theories fibered over an
interval, thereby producing solutions to 4D gauge theories, which likely build
up local $Spin(7)$ backgrounds given by a four-manifold of ADE\ singularities.
In fact, the relevant partially twisted gauge theories have recently been
studied in reference \cite{Heckman:2018mxl} (see also \cite{Gukov:2001hf}).

Along these lines, there are a number of potential physical applications of
the results obtained so far. For one, we note that $G_{2}$ backgrounds lead to
4D $\mathcal{N}=1$ vacua of M-theory and 3D $\mathcal{N}=2$ vacua of type
II\ strings. The physical interpretation of the 2D\ fibration structure also
lends itself to a description in terms of interpolating domain wall
solutions in one higher dimension. It would be very interesting to further
study the properties of these domain walls.

One of the main results of this work is that it is possible to generate
localized matter using T-brane configurations of the fluxed PW system.\ This
in turn suggests a new method for building global M-theory backgrounds with
localized matter in which some of the well-known issues with constructing
codimension seven singularities are simply bypassed. That being said, there do
appear to be deformations which could convert a geometric codimension six
singularity back into a codimension seven singularity. We refer to \cite{barbosa} for a discussion on how such a solution could be engineered from appropriate paths in the
moduli space of the local gauge theory.

In this work we have given an algebraic
characterization of zero modes in local $G_{2}$ backgrounds. That being said,
there are a few outstanding items which would clearly be interesting to
develop further. One concerns the full matter spectrum, even in the case of a
compact three-manifold of ADE singularities in a local $G_{2}$ background. In physical terms, the
matter spectrum is constrained by anomaly cancellation considerations, but at
the moment we have treated each background as a freely adjustable feature of
our models. Presumably this is encoded in the geometry of the spectral
equations for the Higgs field, but it would be interesting to explicitly
verify that this is the case.

Indeed, in the context of intersecting seven-branes in F-theory, spectral
cover methods have been fruitfully applied in developing a streamlined
analysis of many features of localized matter. Since M-theory on a $G_{2}$
background and F-theory on an elliptically fibered Calabi--Yau fourfold both
give rise to 4D $\mathcal{N}=1$ vacua, it is natural to expect an explicit map
which converts the local geometric methods of one description into the other.
The common link in this thread is the appearance of a corresponding
\textquotedblleft local\textquotedblright\ heterotic dual. Developing the
precise form of this match would likely shed light on local methods in
M-theory, F-theory, and heterotic vacua.

It is natural to expect the considerations here to match
on to some macroscopic features as captured by supergravity solutions. Indeed,
there is a class of supergravity solutions known as ``M3-branes'' with a
real codimension four singularity \cite{Cvetic:2001ya, Cvetic:2001zx}. It
would be very interesting to study how these explicit supergravity
solutions match on to features appearing in the PW system.

Finally, we have exclusively focused on the local geometry of
$G_{2}$ backgrounds. We have also seen that at least from the perspective of
the gauge theory, there is no reason to restrict attention to $M$ a rational
homology sphere. This suggests a possible generalization of the
TCS\ construction, as motivated by physical considerations. It is tempting to
ask whether a set of consistency relations can be stated which would allow one
to piece together these local geometries to form a more global
characterization of $G_{2}$ backgrounds with chiral matter.

\section*{Acknowledgements}

We thank S. He, D.R. Morrison, T. Pantev, and S. Sch\"{a}fer-Nameki
for helpful discussions. RB, MC, and JJH thank
the Simons Collaboration workshop on Physics and Special Holonomy held
at the KITP in Santa Barbara for kind hospitality, where part of this
work was completed. The work of MC is supported by DOE Award de-sc0013528y as well as
the Slovenian Research Agency No. P1-0306, and from the
Fay R. and Eugene L. Langberg Chair funds. The work of JJH, CL, and
GZ is supported by NSF CAREER grant PHY-1756996.

\newpage

\appendix

\bibliographystyle{utphys}
\bibliography{Gbranes}

\end{document}